\documentclass[twocolappendix, twocolumn, times]{aastex631}

\newcommand{\Mrange}{\ensuremath{\log M_{\star} \geq 9.66}}
\newcommand{\MrangeS}{\ensuremath{\log M_{\star} \geq 9.66}\ }

\newcommand{\lowmass}{\ensuremath{9.66 \leq \log M_{\star} < 10.5}}

\newcommand{\highmass}{\ensuremath{\log M_{\star} \geq 10.5}}
\newcommand{\highmassS}{\ensuremath{\log M_{\star} \geq 10.5}\ }

\newcommand{\zrange}{\ensuremath{0.1 \leq z \leq 1.0}}
\newcommand{\zrangeS}{\ensuremath{0.1 \leq z \leq 1.0}\ }

\newcommand{\zone}{\ensuremath{0.1 \leq z < 0.35}}
\newcommand{\ztwo}{\ensuremath{0.35 \leq z < 0.5}}
\newcommand{\zthree}{\ensuremath{0.5 \leq z < 0.7}}
\newcommand{\zfour}{\ensuremath{0.7 \leq z \leq 1.0}}

\newcommand{\zoneS}{\ensuremath{0.1 \leq z < 0.35}\ }

\newcommand{\zfourS}{\ensuremath{0.7 \leq z \leq 1.0}\ }

\newcommand{\lhalo}{\ensuremath{L_{{halo}}}}
\newcommand{\lhaloS}{\ensuremath{L_{{halo}}}\ }

\newcommand{\ms}{\ensuremath{M_{\star}}}
\newcommand{\msS}{\ensuremath{M_{\star}}\ }

\newcommand{\sersicS}{S\'ersic }
\newcommand{\angs}{\ensuremath{\text{\normalfont\AA}}\ }
\newcommand{\ang}{\ensuremath{\text{\normalfont\AA}}}

\usepackage{natbib}
\usepackage{hyperref}
\newcolumntype{P}[1]{>{\centering\arraybackslash}p{#1}}
\usepackage{hhline}
\usepackage{makecell}
\usepackage{amsmath}
\usepackage{breqn}
\usepackage{multirow}
\usepackage{ragged2e}
\usepackage{enumitem}
\usepackage[hang,flushmargin]{footmisc}
\usepackage{lipsum} 
\usepackage{afterpage}
\usepackage{xcolor}

\begin{document}

\title{The Role of Cluster Environments in Quiescent Galaxy Stellar Halo Assembly}

\author[0009-0002-1526-0086]{Devin J. Williams}
\affiliation{Institute for Computational Astrophysics and Department of Astronomy \& Physics, Saint Mary's University, 923 Robie Street, Halifax, NS B3H 3C3, Canada}
\email{\textcolor{black}{Email:} devin.williams@smu.ca}

\author[0000-0003-4797-5246]{Ivana Damjanov}
\affiliation{Institute for Computational Astrophysics and Department of Astronomy \& Physics, Saint Mary's University, 923 Robie Street, Halifax, NS B3H 3C3, Canada}

\author[0000-0002-7712-7857]{Marcin Sawicki}
\affiliation{Institute for Computational Astrophysics and Department of Astronomy \& Physics, Saint Mary's University, 923 Robie Street, Halifax, NS B3H 3C3, Canada}

\author[0000-0001-5079-1865]{Harrison Souchereau}
\affiliation{Department of Astronomy, Yale University, New Haven, CT 06511, USA}
\affiliation{Institute for Computational Astrophysics and Department of Astronomy \& Physics, Saint Mary's University, 923 Robie Street, Halifax, NS B3H 3C3, Canada}

\author{Lingjian Chen}
\affiliation{Institute for Computational Astrophysics and Department of Astronomy \& Physics, Saint Mary's University, 923 Robie Street, Halifax, NS B3H 3C3, Canada}

\author[0000-0001-8325-1742]{Guillaume Desprez}
\affiliation{Institute for Computational Astrophysics and Department of Astronomy \& Physics, Saint Mary's University, 923 Robie Street, Halifax, NS B3H 3C3, Canada}

\author[0009-0009-2742-7702]{Angelo George}
\affiliation{Institute for Astronomy and Astrophysics, Academia Sinica, No.1, Sec. 4, Roosevelt Rd, Taipei 106319, Taiwan}

\author[0000-0001-8221-8406]{Stephen Gwyn}
\affiliation{NRC Herzberg Astronomy and Astrophysics, 5071 West Saanich Road, Victoria, BC V9E 2E7, Canada}

\author{St\'ephane Arnouts}
\affiliation{Aix Marseille Universit\'e, CNRS, CNES, LAM, 38 rue Fr\'ed\'eric Joliot-Curie, Marseille 13388 cedex 13, France}

\begin{abstract}

External interactions drive galaxy stellar mass growth and morphological evolution. As stellar haloes—assembled largely via hierarchical accretion—preserve signatures of these processes, their growth probes how environment regulates galaxy evolution. We investigate how cluster environments influence quiescent galaxy (QG) stellar halo assembly over \zrangeS in a sample of 2,168 cluster and 94,479 field QGs of \Mrange. Extended emission is traced via rest-frame $g$-band surface brightness ($\mu_g$) profiles extracted from deep HSC-SSP $grizy$ imaging. We study stellar halo assembly trends by linking median $\mu_g$ profile evolution to the underlying mass growth in galaxy subpopulations. Over \zrange, cluster QGs build up stellar haloes faster than field QGs, with a $\sim23\%$ and $\sim40\%$ larger increase in integrated stellar halo luminosity (\lhalo) in the low-mass (\lowmass) and high-mass (\highmass) samples, respectively. High-mass cluster QGs host more luminous stellar haloes than the field (mean cluster-to-field \lhaloS ratio of $\sim1.2$), while low-mass cluster QGs host less luminous stellar haloes than their field counterparts (mean ratio of $\sim0.87$). Among cluster QGs of $\log M_{\star} \geq 10$, \lhaloS increases with host cluster mass, but decreases for cluster QGs of $\log M_{\star} < 10$. These results suggest higher-mass cluster QGs ($\log M_{\star} \geq 10$) experience enhanced stellar halo growth over \zrangeS fueled by increased merger-driven accretion, likely from minor mergers in cluster outskirts or in pre-infall group and filament environments. Lower-mass cluster QGs ($9.66 \leq \log M_{\star} < 10$) instead have suppressed stellar halo growth in clusters and likely lose outer stellar material to environmental stripping or accretion by high-mass galaxies during mergers.

\end{abstract}


\section{Introduction}\label{main-sec:intro}

Two broad classes of processes govern stellar mass growth and morphological transformations in galaxies over cosmic timescales: internal (secular) evolution and external (environment-driven) interactions (e.g., \citealt{Hickson1997_nature, Lucia2019_nature, Shi2024_nature}). Internal processes include the smooth accretion and regulation of gas within galaxies, which fuels in situ star formation and drives structural evolution through disk instabilities and feedback from active galactic nuclei (AGN) and supernovae (e.g., \citealt{Bergin2007-AA, McKee2007-AA, Fabian2012_AA}). After quenching (see e.g., \citealt{Belli(2018)} for a list of quenching mechanisms), galaxies continue to evolve passively through the aging of stellar populations and secular processes (e.g., disk heating, bar dynamics, and pseudobulge growth). These processes further change galaxy properties (e.g., color, size, and morphology; \citealt{Kormendy1993-bulge, Bouwens1996-passiveEVO, Franx2000-passiveEVO, Merrifield2001-diskheat, Kormendy2004, Geron2024-bar}). External processes include galaxy mergers and environmental mechanisms that redistribute or remove baryonic material when galaxies interact with a deep gravitational potential well or hot intergalactic medium (IGM, \citealt{White1987, White1991, Boylan-Kolchin-mergers, Blanton2009, Peng-2010}). The efficiency and relative importance of these external processes depend strongly on galaxy environmental properties, including the host halo mass and local galaxy density.

Galaxy clusters, the densest environments, host hundreds of galaxies within a deep gravitational potential filled with a hot, dense intracluster medium (ICM; \citealt{Kravtsov2012_AA_ICM_clusters} and references therein). As galaxies interact with the ICM, hydrodynamical processes such as ram-pressure stripping (RPS; \citealt{Abadi1999_RPS, Boselli2006, Boselli-RPS}) and gas supply cut-off (e.g., \citealt{Larson1980, Peng2015_strangulation, Kuutma2017, Brown2017, Trussler2020}) suppress star formation by removing or heating a galaxy’s gas supply without significantly disturbing its existing stellar populations. Conversely, gravitational processes - such as tidal stripping (e.g., \citealt{Merritt1983, Merritt1985, Read2006, Fang2016}) and fly-bys (e.g., \citealt{Moore1996, Moore-1998-harassment, Bialas2015}) - can remove both gas and stars from a galaxy, leading to quenching or the formation of tidal features. These stripped stars contribute to the buildup of intracluster light (ICL)—diffuse light from intergalactic stars within cluster environments—over time (e.g., \citealt{Contini2021_ICL_AA, Montes2022_ICL, Golden-Marx2023, Golden-Marx2025}). Together, these environmental processes drive cluster galaxies to be redder, less star-forming, and exhibit early-type galaxy (ETG) morphologies (e.g., \citealt{Dressler1984, Dokkum-2001-PB, Aguerri2004, vanderwel_2010, Rodriguez2024, Li2025_referenced_me}).

Among external processes, galaxy mergers play a central role in driving galaxy stellar mass assembly and morphological evolution. Major mergers contribute substantial amounts of ex situ stellar material to both the inner and outer regions of the host galaxy, often inducing significant structural transformations into dispersion-dominated elliptical morphologies. By contrast, minor mergers primarily deposit material in galaxy outskirts and produce more subtle morphological changes (e.g., \citealt{Lambas(2012), Hilz2013, Ownsworth(2014), Zhu2022, Jackson(2022)}). The impact of mergers also depends on gas content, with gas-rich (wet) mergers capable of triggering bursts of both star formation and AGN activity. In contrast, gas-poor (dry) mergers predominantly add stellar mass without increasing the cold gas supply (e.g., \citealt{Bell-drymergers, Trujillo(2011), Ellison2013, Ellison2020, Ellison2025_merger_SFR, Li2023-starburst}). Observations of massive ($\log M_{\star} \geq 11$, where \msS denotes stellar mass in solar units, $M_{\star}=M_{\star}/M_{\odot}$) quiescent galaxies (QGs) and ETGs over a broad redshift range ($0 < z < 2$) indicate that dry major mergers dominate their stellar mass assembly (e.g., \citealt{Bell-drymergers, vanDokkum(2010), Bernardi2011}). Additionally, the accelerated size growth observed in QGs over time (e.g., a $2.5{-}4\times$ increase in size since $z = 1{-}2$) has been attributed to dry minor mergers, which are efficient at increasing galaxy sizes (e.g. $R_e \propto M^{2{-}2.5}$; \citealt{Bezanson(2009), Naab(2009)}) without significantly enhancing star formation (e.g., \citealt{Daddi2005-sizeevo, Trujillo(2012), vanderWel(2014), Damjanov(2019), Angelo2024}).

High velocity dispersions in cluster environments, particularly in the centers, are thought to suppress mergers between cluster galaxies due to their large relative velocities (e.g., \citealt{Omori2023, Sureshkumar2024, Yoon2024}). Despite this, some low-redshift studies ($z \lesssim 0.15$) find evidence of mergers in clusters in the form of tidal (post-merger) features and close companions (e.g., \citealt{Iodice2017, Ribeiro2023_cluster_mergers, Edwards2024_cluster_mergers, Kim2024_cluster_mergers, HyeongHan2025}). At higher redshifts ($0.75<z<1.2$), \cite{Lin2010} found that dry mergers preferentially occur in high-density environments. Accelerated size growth observed in cluster ETGs has been attributed to increased minor merger-driven accretion (e.g., \citealt{Noordeh2021, Afanasiev2023}), although other studies find that galaxy size growth is independent of environment (e.g., \citealt{Matharu2020_cluster_QG_sizes, Figueira2024_cluster_QG_sizes, angelo_p2_final}). These mixed results highlight that a comprehensive understanding of galactic evolution in the densest environments remains challenging. At $z\leq 1$, cluster populations are dominated by quenched galaxies, making QGs ideal targets for studying how dense environments influence galaxy assembly (e.g., \citealt{Dressler1984, van_der_burg_2018, Wang2022, Brown2023}).

Galaxy outskirts provide a particularly sensitive probe of the environmental effects, as long dynamical timescales at large radii preserve signatures of past interactions (e.g., \citealt{Bell1995-shellstream, Delgado2008-shellstream, Delgado2015-shellstream, Iodice2017, Dey2023-shellstream}). Furthermore, galaxy stellar haloes - extended distributions of stars and gas surrounding galaxies - are predicted to form primarily through hierarchical assembly via successive accretion events (e.g., \citealt{Cooper2010, Pillepich2014, Cook(2016)}). Observations tracing stellar halo evolution therefore provide a direct probe of how environmental interactions shape QG mass assembly over time.

Extended stellar halo emission is commonly studied using galaxy surface brightness ($\mu$) profiles (or \emph{light} profiles; e.g., \citealt{Dsouza2014, Buitrago(2017), Wang2019-profiles, Spavone2020, Spavone2022, Gilhuly(2022)}), which cosmological simulations identify as tracers of galaxy assembly (e.g., \citealt{Hopkins2010-sizemass, Hilz2013, Hirschmann(2015), Cook(2016)}). By integrating these profiles and assuming stellar mass-to-light ratios ($M_{\star}/L$), stellar halo mass fractions can be estimated and used as proxies for accreted (ex situ) stellar mass (e.g., \citealt{Elias2018, Huang(2018), Merritt2020}). A study of stellar mass density profiles (derived from light profiles) of massive QGs ($\log M_{\star}>11.4$) at $0.3<z<0.5$ in HSC-SSP data shows that the stellar mass fraction beyond 10 kpc (i.e., the stellar halo region) agrees with predicted ex situ fractions from cosmological simulations \citep{Huang(2018)}. In \cite{Huang2018-DM}, the authors demonstrate that massive QGs residing in more massive dark matter haloes exhibit more prominent outer envelopes in their mass profiles, suggesting larger stellar haloes are found in denser environments. Based on a small sample of ETGs ($\log M_{\star} \sim 9{-}11$) in the Fornax cluster (i.e., $z\sim 0$), more massive ETGs host larger stellar halo light fractions in their light profiles and preferentially occupy denser regions of the cluster \citep{Spavone2020}.

In our previous work (\citealt{DevinP1}, hereafter DJW2025), we studied stellar halo assembly in the field over $0.2 \leq z \leq 1.1$ in a sample of 330,877 QGs and star-forming galaxies (SFGs) from the HSC-SSP and CLAUDS surveys (Section \ref{main-sec:data}), constraining the relative contributions from in situ star formation and ex situ accretion. Following the methodology developed in DJW2025 for extracting and analyzing galaxy light profiles, we now examine how cluster environments influence stellar halo assembly in QGs. Our mass-complete sample (\Mrange) includes 94,479 field QGs and 48 galaxy clusters with 2,168 cluster member QGs. Our analysis thus enables a population-level investigation of the net effect of all external processes at play in dense environments on galaxy stellar halo assembly.

This paper is organized as follows. Section \ref{main-sec:data} details the photometric datasets we use and the selection of our cluster and field QG samples. Section \ref{main-sec:methods} outlines our methodology for extracting galaxy rest-frame $g$-band light profiles and for quantifying the redshift evolution in the median profiles of different galaxy subsamples. We present our main results in Section \ref{main-sec:results}, with further discussion in Section \ref{main-sec:discussion}. We summarize our main conclusions in Section \ref{main-sec:conclusions}. Throughout this work, magnitudes are quoted in the AB system and a $\Lambda$CDM cosmological model with $\Omega_M$ = 0.3, $\Omega_{\Lambda}$ = 0.7, and $H_0$ = 70 km s$^{-1}$Mpc$^{-1}$ is assumed.

\section{Datasets and Galaxy Sample Selection}\label{main-sec:data}

\subsection{Photometric Data and Galaxy Catalogs}\label{sec:data_catalogues}
For this work, we use broadband $grizy$ imaging ($\sim 4000-10000\ang$) from the third Public Data Release (PDR3\footnote{PDR3 Data from HSC-SSP can be obtained at: \url{https://hsc-release.mtk.nao.ac.jp/doc/index.php/data-access__pdr3/}.}) of the Hyper Suprime-Cam Subaru Strategic Program (HSC-SSP, \citealt{Aihara(2022)}). The data are drawn from the Deep (26 deg$^2$) and UltraDeep (3.5 deg$^2$, embedded within Deep) layers, which together span four widely separated extragalactic fields (XMM-LSS, E-COSMOS, ELAIS-N1, and DEEP2-3). We select the images produced by the HSC-SSP pipeline (\texttt{hscPipe}, \citealt{Bosch(2018)}) with the global sky subtraction applied, as it has been shown to better preserve the faint emission in galaxy outskirts \citep{Aihara2019-PDR2, Aihara(2022)}. In Appendix \ref{appendixB}, we show how light profile extractions are affected when using either the global or local sky-subtracted HSC-SSP images.

We select galaxies from the CLAUDS+HSC-SSP photometric catalogs of \citealt{Desprez2023} (\texttt{hscPipe}/\texttt{Phosphoros} versions) which cover a combined $\sim$18 deg$^2$ over the four Deep+UltraDeep fields. Additional data beyond HSC-SSP used in the construction of these catalogs include $U$-band imaging from the CFHT Large Area $U$-band Deep Survey (CLAUDS; \citealt{Sawicki(2019)}) and VIRCAM near-IR wavelength coverage ($Y$, $J$, $H$, and $K_s$ bands) from the VIDEO \citep{Jarvis2013-VIDEO} and UltraVISTA surveys \citep{UltraVISTA}. From these CLAUDS+HSC-SSP catalogs, we obtain galaxy coordinates, magnitudes derived from \texttt{hscPipe cmodel} fluxes corrected for Galactic extinction, and photometric redshifts ($z_{phot}$) computed via spectral energy distribution (SED) fitting with \texttt{Phosphoros} using $Ugrizy$ bands (or $Ugrizy+JHK_s$ when available). Galaxies are not required to have full multi-wavelength coverage; SED fitting is performed using all available bands for each source. Average errors on $z_{phot}$ are $\sigma_{(1+z)} \sim 0.04$, but both precision and outlier fractions ($\eta$) worsen with magnitude ($\sigma \sim 0.03$ at $m_i \leq 22.5$ and $\sim 0.09$ at $m_i \leq 26$; $\eta$ increases from $\sim 4\%$ to $\sim 29\%$; \citealt{Desprez2023}). 

We obtain updated stellar masses for CLAUDS+HSC-SSP galaxies derived via \texttt{LePhare} SED fitting by \cite{LJ_PHD}. The authors combine $Ugrizy$ photometry with $JHKs$ in the XMM-LSS field and additional IRAC 3.6$\mu$m and 4.5$\mu$m data from the SHIRAZ Survey \citep{spitzer_mass_data} in the remaining fields. The addition of IR data improves stellar mass estimates by more effectively tracing the mass-dominant older stellar populations in galaxies and constraining the long-wavelength dust emission associated with star formation (e.g., \citealt{Zibetti2009, Wen2013, Kouroumpatzakis2023}). In these runs, redshifts are fixed to the $z_{phot}$ from \cite{Desprez2023}. Galaxy models are constructed from SED templates of \cite{Bruzual-stellar-pop} assuming a \cite{Chabrier2003} initial mass function (IMF) and exponentially decaying star-formation histories (SFR $\propto e^{- \tau}$ with $\tau = 0.01{-}3$ Gyr). Our stellar masses show strong agreement with the COSMOS2020 catalog \citep{COSMOS2020}, with median offsets ($\Delta \log M_{\star}$) ranging from $\sim$0.01 dex at lower masses to $\sim$0.12 dex at the high-mass end ($\log M_{\star} \sim 11.5$; \citealt{angelo_p2_final}). Individual stellar mass uncertainties are characterized by the 68\% range of the stellar mass probability distributions produced by \texttt{LePhare}, with typical $1\sigma$ uncertainties of $\sim0.11$ dex for low-mass galaxies (\lowmass) and $\sim0.09$ dex for high-mass galaxies (\highmass).

\subsection{Initial Sample and Quiescent Galaxy Classification}\label{sec:data_quality_cuts_QG_selection}
We apply the same quality cuts to the initial sample of galaxies drawn from the CLAUDS+HSC-SSP catalogs as in DJW2025 (Table 1 in \citealt{DevinP1}). In short, we limit galaxies to $m_i\leq 25$ AB as uncertainties on $z_{phot}$ grow larger at fainter magnitudes \citep{Desprez2023}. We remove point source objects misclassified as galaxies by applying \texttt{isStar}=False, \texttt{isCompact}=False, and \texttt{i_compact_flag}=False. We remove galaxies with image defects such as satellite trails or nearby bright star masks by setting \texttt{isOutsideMask}=1. We remove galaxies with unreliable photometry measurements due to failed \texttt{cmodel} magnitudes by setting \texttt{CMODEL\_FAIL\_FLAG}$<2$. We omit a small number of galaxies ($\sim 2\%$) with failed light profile extractions caused by image artifacts not caught by the quality cuts or issues from the source masking procedure implemented during profile extraction (Section \ref{sec:methods_image_corections_profile_extractions}).

We limit galaxy stellar masses to \MrangeS based on \msS completeness in the CLAUDS+HSC-SSP datasets from the study of \cite{LJ_PHD}, who used the empirical method from \cite{Pozzetti2010} to calculate the \msS completeness limit as a function of redshift. We restrict the redshift range of our sample to \zrangeS (Section \ref{sec:data_final_sample_selection}).

Galaxies in the CLAUDS+HSC-SSP catalogs are classified as quiescent via a modified $NUVrK$ color-color selection \citep{Arnouts(2013), Moutard(2018), Moutard(2020), angelo_p2_final}. In this method, any dependence on stellar mass (\ms) and redshift ($z$) in selecting QGs is accounted for by computing an optimal separation boundary parametrized as:

\begin{equation}
\begin{split}
    NUV-r &> (c_{1}z + c_{2}\log M_{\star} + c_{3})[(r-K) + c_{4}] \\
    NUV-r &> c_{5}.
\end{split}
\label{eq:NUVRK}
\end{equation}

The constants ($c_{1}{-}c_{5}$) in Equation \ref{eq:NUVRK} are determined empirically by calibrating with the COSMOS2020 catalog from \cite{COSMOS2020}, where galaxies with $\log$sSFR $<-11$ were taken to be quiescent. Equation \ref{eq:NUVRK} is optimized to reduce contamination from SFGs while increasing completeness in the QG sample \citep{angelo_p2_final}.

\begin{figure}[ht]
\centering
\includegraphics[width=0.47\textwidth]{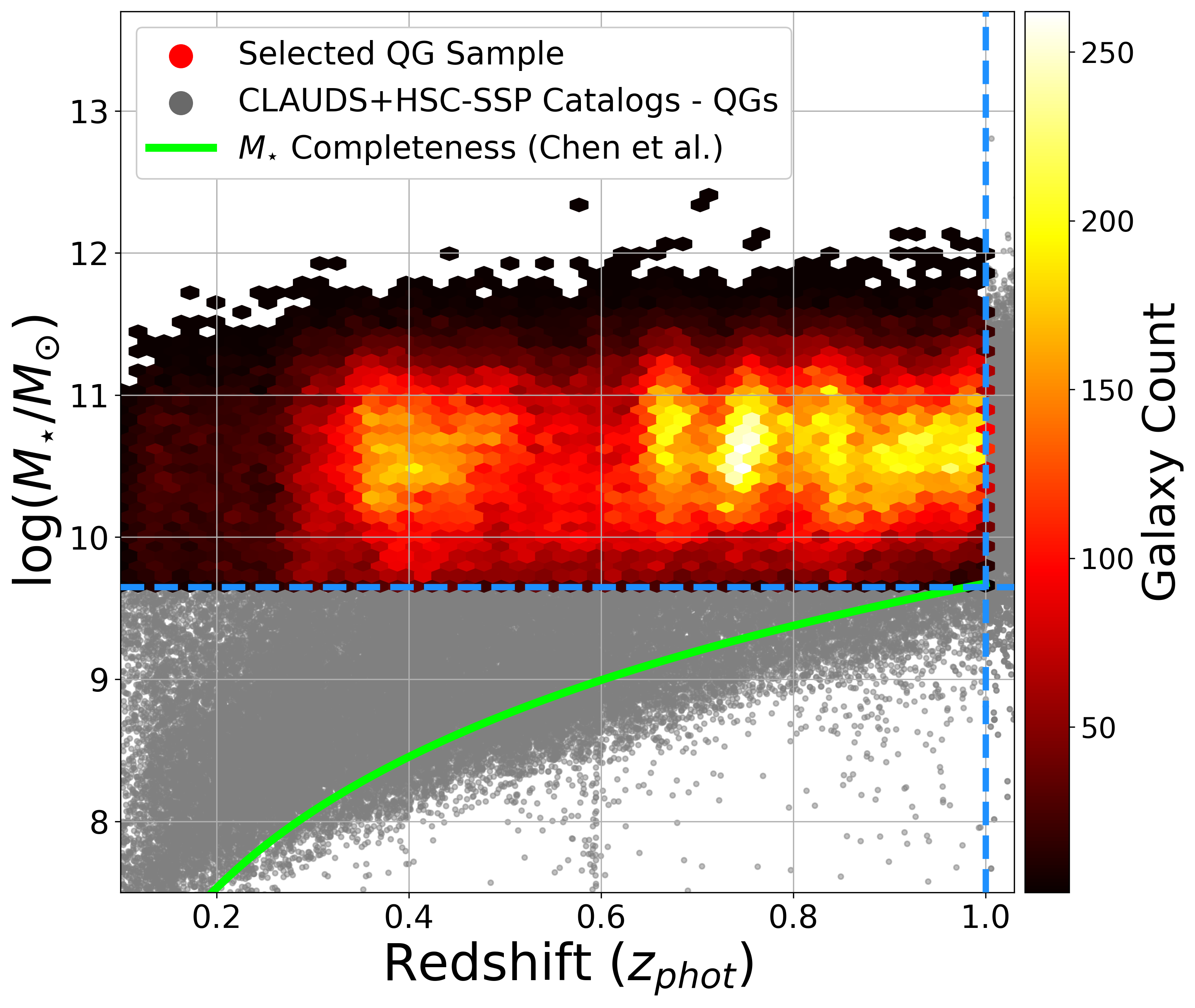}
\caption{\footnotesize{Stellar mass as a function of redshift for our initial selected QG sample (colored density points) and the full parent sample of QGs from the CLAUDS+HSC-SSP photometric catalogs (grey points). The blue horizontal line at $\log M_{\star}=9.66$ represents our stellar mass limit based on \msS completeness from \citealt{LJ_PHD} (green curve). The blue vertical line at $z=1.0$ denotes our upper redshift limit.}
\label{fig:DATA_MZ_parent}}
\end{figure}

In summary, our initial mass-complete sample comprises 102,700 QGs of \MrangeS at \zrangeS across the combined four deep+UD fields (Section \ref{sec:data_catalogues}). Figure \ref{fig:DATA_MZ_parent} shows the stellar mass versus redshift diagram for our selected sample (colored points) compared to the full parent sample (grey points) available in the CLAUDS+HSC-SSP catalogs. We describe additional minor reductions to this sample ($\sim2\%$) in Section \ref{sec:methods_dense_enviro_tests}, based on tests of the performance of our light profile extraction procedure in dense environments. 

\subsection{Identifying Cluster Members}\label{sec:data_cluster_selection}

\begin{figure*}
\centering
\includegraphics[width=0.99\textwidth]{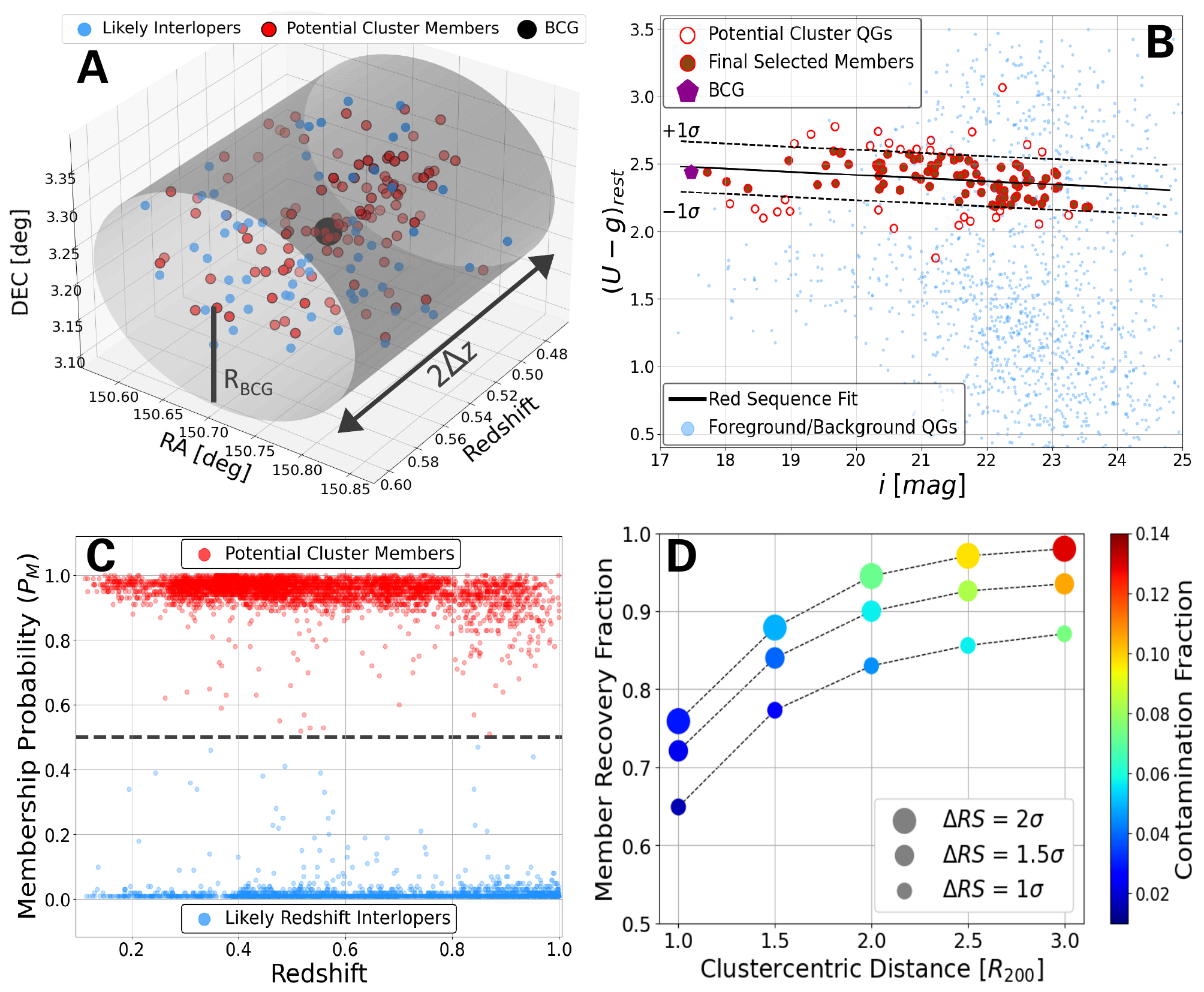}
\caption{\footnotesize Demonstration of key steps to our cluster member-finding procedure. Panel A: Cylindrical volume with a comoving radius ($R_{BCG}$) of 3 cMpc and length of $2\Delta z$ where $\Delta z = 0.05(1+z)$ centered on the BCG (black dot). Red dots highlight potential cluster members with high membership probability ($P_M \geq 0.5$), while blue dots represent likely redshift interlopers. Panel B: Color-magnitude diagram used to fit a linear red sequence relation (black line) to each cluster. Potential cluster galaxies are shown as red open circles, with final selected members filled in with brown. Foreground/background field QGs are shown as small blue dots. Panel C: Cluster membership probabilities ($P_M$) vs. redshift for all CLAUDS+HSC-SSP galaxies identified by our cluster member-finding algorithm after $N=100$ iterations. The black horizontal line at $P_M =0.5$ denotes our separation between selected and rejected cluster membership. Panel D: Results of testing our procedure on HectoMAP cluster and field galaxy samples. The y-axis represents the member recovery fraction for a given set of parameters, while the dot colors represent the contamination fraction from interlopers. The x-axis shows different limits on allowed clustercentric distances ($R_{BCG}/R_{200}$), while dot sizes represent different allowed color offsets from a cluster's red sequence.
\label{fig:data_cluster_code_4panel}}
\end{figure*}
We implement our own cluster member finding algorithm to identify cluster galaxies in our CLAUDS+HSC-SSP sample rather than relying on existing photometric cluster member catalogs available for the HSC-SSP survey region (e.g., \citealt{Oguri2018, Wen2021_cluster_catalog_1pz}). Our approach combines two commonly used techniques: an initial selection of galaxies near cluster centers in projected position and redshift space, followed by refinement using the red sequence method (e.g., \citealt{Gladders2000, Gladders2005, redmapper_2014, Oguri2014}). Unlike previous catalogs in our survey region, however, galaxy properties in the CLAUDS+HSC-SSP catalog \citep{Desprez2023} are derived from SED fitting that includes $U$-band and NIR flux measurements. This additional data yields more robust stellar mass estimates (Section \ref{sec:data_catalogues}), and most importantly, more accurate photometric redshifts out to $z\sim 0.75$ (e.g., $\sim 50\%$ decrease in the outlier fraction $(\eta$: $\Delta z > (1+z_{spec})\times 0.15)$ and $\sim 30\%$ decrease in the scatter $(\sigma = 1.48 \times \mathrm{median}\!\left(\frac{\Delta z}{1+z_{\rm spec}}\right))$, where $\Delta z = |z_{spec} - z_{phot}|$; \citealt{Sawicki(2019)}), which provides a more robust identification of cluster members by better constraining their proximity to cluster centers in redshift space. Here, we briefly describe our procedure for identifying quiescent cluster galaxies (QCGs) and demonstrate the main steps in Figure \ref{fig:data_cluster_code_4panel}.

First, we obtain locations of clusters using the coordinates of spectroscopically confirmed brightest cluster galaxies (BCGs) from the catalogue of \cite{Oguri2018}, and use the BCG position to define the cluster center. For each cluster, we search for CLAUDS+HSC-SSP galaxies that lie within a cylindrical volume centered on the BCG with a comoving radius of 3 cMpc and length of $2\Delta z$ where $\Delta z = 0.05(1+z)$. Panel A in Figure \ref{fig:data_cluster_code_4panel} illustrates this step for a BCG (black dot) at $z_{spec} = 0.53598$ in the COSMOS field. 

Next, we refine the selection of potential QCGs using the red sequence method, which leverages the fact that cluster QGs at the same redshift exhibit similar colors \citep{Gladders2005, MaxBCG_2007, redmapper_2014}. For each cluster, we fit a linear red sequence relation (RS; black solid line in panel B of Figure \ref{fig:data_cluster_code_4panel}) to the $U-g$ (rest-frame at the cluster redshift) versus $i$-band color–magnitude data of the potential members. Rest-frame $U-g$ colors are computed from observed-band apparent magnitudes ($grizy$), where the pair of filters used for each galaxy is selected to best trace rest-frame $U$ ($3000$\angs) and $g$ ($5000$\angs) emission following $\lambda_{\rm rest} = \lambda_{\rm obs}(1+z)^{-1}$, where $\lambda_g$ and $\lambda_{obs}$ represent rest-frame $g$-band and observed-band wavelengths, respectively. We identify galaxies that lie near the cluster red sequence ($\Delta RS \leq 1\sigma$; Section \ref{sec:data_final_sample_selection}) as final cluster members (brown-filled dots in panel B of Figure \ref{fig:data_cluster_code_4panel}).

We perform our cluster member finding algorithm iteratively ($N=100$) to account for uncertainties in galaxy $z_{phot}$. We draw 100 $z_{phot}$ for each CLAUDS+HSC-SSP galaxy from their redshift probability distribution functions (PDFs) and run the entire member finding procedure for all draws independently. This yields a membership probability ($P_M$) for each galaxy, based on the fraction of iterations in which it is confirmed to be a cluster member. We adopt $P_M\geq0.5$ when selecting the final set of cluster members (Section \ref{sec:data_final_sample_selection}) to reduce contamination from redshift interlopers, similar to previous studies (e.g., \citealt{George2011_clustercode, Rozo2015_clustercode, Sohn2021_hectomap_all, angelo_p2_final}). Panel C in Figure \ref{fig:data_cluster_code_4panel} shows how this cut of $P_M\geq 0.5$ eliminates the vast majority of likely redshift interlopers with low $P_M$ (blue dots) while maintaining high recovery of likely members with high $P_M$ (red dots).

To validate the performance of our cluster member finding procedure, we test it using the HectoMAP cluster and field galaxy catalogs of \cite{Sohn2021_hectomap_cluster, Sohn2021_hectomap_all}. The HectoMAP survey lies within the HSC-SSP Wide footprint. It provides a large sample of spectroscopically confirmed cluster members, offering a reliable ground truth cluster sample which we can use to assess the performance of our member-finding procedure. In total, we obtain 346 galaxy clusters and 4,897 member galaxies with $D_n4000 >1.5$ (likely to be quiescent, e.g., \citealt{Damjanov(2019)}), and an additional 16,261 field galaxies confirmed not to be part of a group or cluster. In these tests, we use average $z_{phot}$ errors ($\sigma _z \sim 0.03$) from our CLAUDS+HSC-SSP sample to adjust the individual $z_{spec}$ of HectoMAP galaxies in each iteration to make the tests more comparable to our photometric sample. Varying the red sequence width ($\Delta RS$) and normalized clustercentric distance ($R_{BGC}/R_{200}$) affects both the member recovery rate and contamination fraction ($N_{field}/N_{true\ mem}$). Panel D of Figure \ref{fig:data_cluster_code_4panel} shows how these fractions change with different parameter choices. We discuss our final chosen parameter cuts in Section \ref{sec:data_final_sample_selection}.

\subsection{Final Cluster and Field Galaxy Subsamples}\label{sec:data_final_sample_selection}

A total of 5,464 QGs are identified as potential cluster galaxies from implementing our iterative cluster member finding algorithm (Section \ref{sec:data_cluster_selection}) on the CLAUDS+HSC-SSP sample. As this initial sample contains many likely field interlopers (e.g., it contains the blue points with low $P_{M}$ in panel C in Figure \ref{fig:data_cluster_code_4panel}), we restrict the final cluster QG sample to galaxies with a membership probability of $P_M \geq 0.5$ (this cut removes 1,745 galaxies). Additionally, we limit cluster galaxies to being within 2$R_{200}$ of their host cluster (removing 869 galaxies in total) and within $\Delta RS=1 \sigma$ of their cluster's fitted red sequence (removing an additional 682 galaxies in total). Based on the tests summarized in panel D of Figure \ref{fig:data_cluster_code_4panel}, our choice of parameters likely achieves high cluster member completion ($\gtrsim 84\%$) while maintaining a very low contamination fraction ($\lesssim 4\%$), which ensures a more purified cluster sample. Our choice of 2$R_{200}$ as the outer boundary is motivated by the work of \cite{Pizzardo2024}, who show that $R_{\nu_{min}}$—the radius where radial velocity reaches its minimum and the cluster infall motion is strongest—typically occurs at $\sim$ 2-3$R_{200}$ in simulated galaxy clusters across our redshift range. Adopting 2$R_{200}$ therefore ensures that all selected cluster galaxies lie well within the region influenced by the cluster potential and are subject to its environmental effects.

\begin{figure}[h]
\centering
\includegraphics[width=0.47\textwidth]{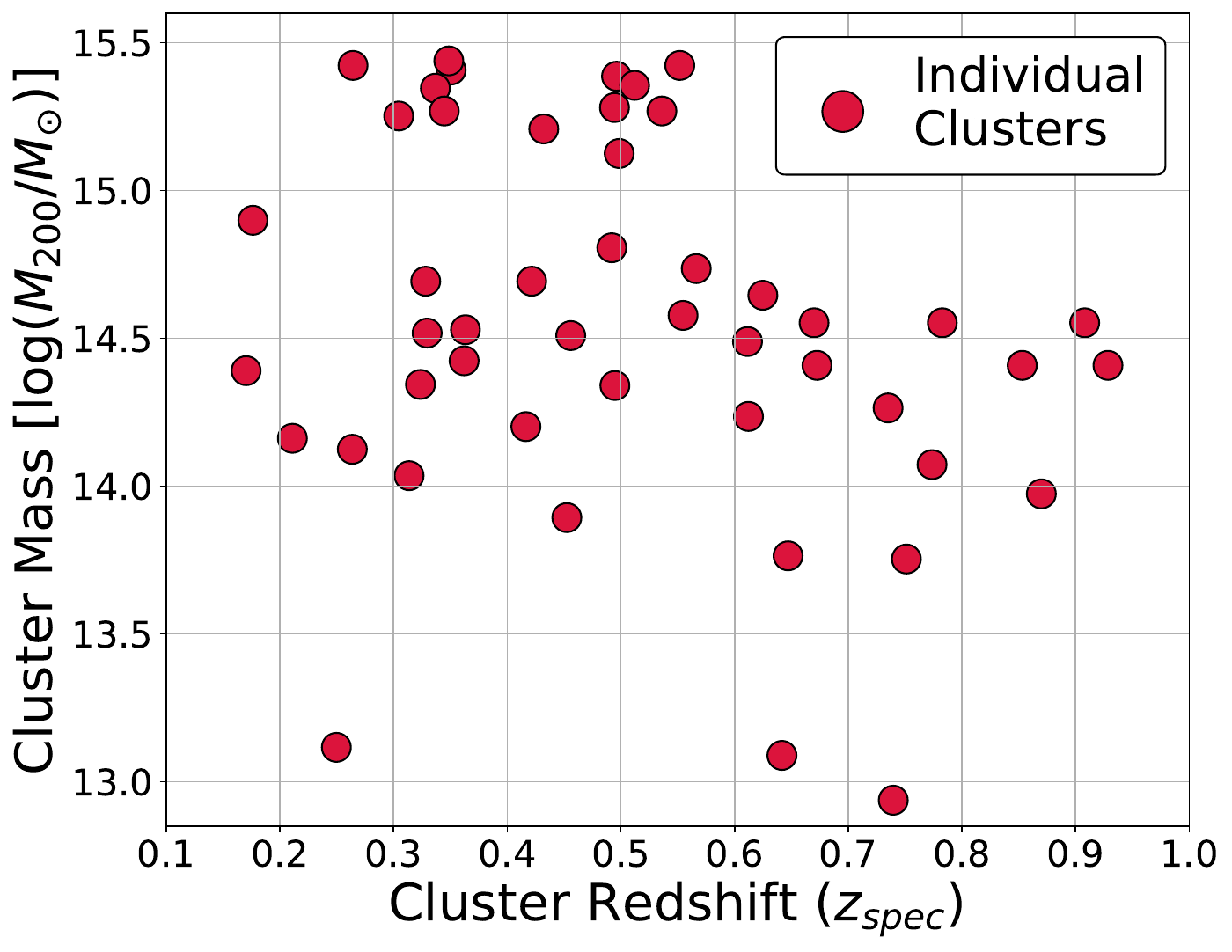}
\caption{\footnotesize{Cluster mass ($M_{200}$) as a function of cluster spectroscopic redshift for all 48 clusters in our final sample. Cluster masses represent dark matter halo masses estimated from the relation between DM halo mass and BCG stellar mass from \cite{Leauthaud2012_BCG_halomass}.}
\label{fig:DATA_cluster_NZ_plot}}
\end{figure}

Our final cluster sample comprises 48 galaxy clusters with 2,168 QG members. We remove the BCGs from our sample as they represent unique objects that differ from the cluster member galaxy population and are beyond the scope of this study. However, we obtain a cluster dark matter (DM) halo mass ($M_{200}$) for each cluster using the relation between DM halo mass and BCG stellar mass from \cite{Leauthaud2012_BCG_halomass}. We use these DM halo masses as proxies for cluster mass in Section \ref{sec:results_M200} to probe the impact of host cluster mass on QG stellar haloes. Figure \ref{fig:DATA_cluster_NZ_plot} shows these estimated cluster masses as a function of cluster redshift ($z_{spec}$) for our 48 clusters.

\begin{deluxetable*}{lccccc}
\tablenum{1}
\tablecaption{Number of field and cluster QGs in each \msS and $z$ bin combination (i.e., subsample) we define.}
\label{table:MZ_table}
\tablehead{
  \colhead{Redshift Bins:} & \colhead{$0.1\leq z < 0.35$} & \colhead{$0.35\leq z < 0.5$} &  \colhead{$0.5\leq z < 0.7$} & \colhead{$0.7\leq z \leq 1.0$} & \colhead{\textbf{Total}}}
\startdata
\multicolumn{6}{c}{\textbf{Low-Mass Galaxies (\lowmass)}} \\
Field QGs   & 4967 & 8674 & 10608 & 18800 & \textbf{43049} \\
Cluster QGs & 185 & 462 & 261 & 130 & \textbf{1038} \\
\multicolumn{6}{c}{\textbf{High-Mass Galaxies (\highmass)}} \\
Field QGs   & 4071 & 9881 & 11359 & 26119 & \textbf{51430} \\
Cluster QGs & 195 & 506 & 232 & 197 & \textbf{1130} \\
\enddata
\end{deluxetable*}

Our field sample consists of all other QGs that passed the quality cuts (Section \ref{sec:data_quality_cuts_QG_selection}) and are within \MrangeS and \zrange. To increase the purity of the field sample, we remove the entire sample of initially identified potential cluster QGs (i.e., the 5,464 galaxies with $P_M> 0$), and not just those selected in the final purified cluster QG sample. Following this purification step, 94,479 galaxies remain in our final field QG sample. 

As an additional verification that we are selecting true cluster member galaxies, we compared the distributions of star formation rates (SFR), specific-star formation rates (sSFR), rest-frame $U{-}g$ colors, and $NUV$-band absolute magnitudes between our final field and cluster QG samples. It has been well established in the literature that galaxies in denser cosmic environments tend to be less star-forming and redder in colour (e.g., \citealt{Moore-1998-harassment, Peng-2010, Hahn2015, Gu2021, Wang2022}). As expected, our cluster QG sample exhibits lower mean SFRs and sSFRS (-1 dex and -1.6 dex, respectively), redder colours (+0.3 dex), and fainter $NUV$ magnitudes (+0.5 dex) than the field QG sample.

To study how different galaxy populations are building up their stellar haloes over time, we divide our cluster and field QG samples into smaller bins of \msS and redshift (Table \ref{table:MZ_table}). As in DJW2025, we divide the galaxy sample into low-mass (\lowmass) and high-mass (\highmass) galaxies, motivated by the pivot mass ($\log (M_p/ M_{\odot}) \sim 10.5 \pm 0.3$) of observed galaxy size–stellar mass relations (e.g., \citealt{Lange-pivot, Mowla-pivot, Kawin2021, Damjanov2022-pivot-PB, Angelo2024}). This pivot point marks the transition into a steeper size-stellar mass relation slope for more massive galaxies. This change in slope has been interpreted as an increased contribution from merger-driven accretion in more-massive galaxies, based on predicted ex situ fractions of galaxies in cosmological simulations (e.g., \citealt{Rodriguez(2016), Tacchella2019, Davison2020, Husko(2022)}). We note that varying the cutoff between low- and high-mass galaxies across $\log M_{\star} \sim 10.5 \pm 0.3$ does not affect the overall trends or conclusions presented in this work (Section \ref{main-sec:results}).

Our four redshift bins (Table \ref{table:MZ_table}) loosely follow those of DJW2025, but are modified slightly due to the smaller size of the cluster QG sample and the discrete redshifts of clusters. The lowest and most sparsely populated redshift bin, \zone, spans the $\sim 2.5$ Gyr of cosmic time, and the remaining three redshift bins cover similar cosmic time intervals ($\sim 1.2-1.4$ Gyr).

In summary, our final CLAUDS+HSC-SSP galaxy sample includes 94,479 field QGs and 2,168 cluster QGs with \MrangeS spread over a redshift range of \zrange. Table \ref{table:MZ_table} provides the number of galaxies in each subsample we define (i.e. \msS+ $z$ bin combinations).

\section{Extracting Galaxy Light Profiles and Analyzing Median Evolution}\label{main-sec:methods}
\subsection{Image Corrections and Individual Profile Extractions}\label{sec:methods_image_corections_profile_extractions}
\begin{figure*}
\centering
\includegraphics[width=0.96\textwidth]{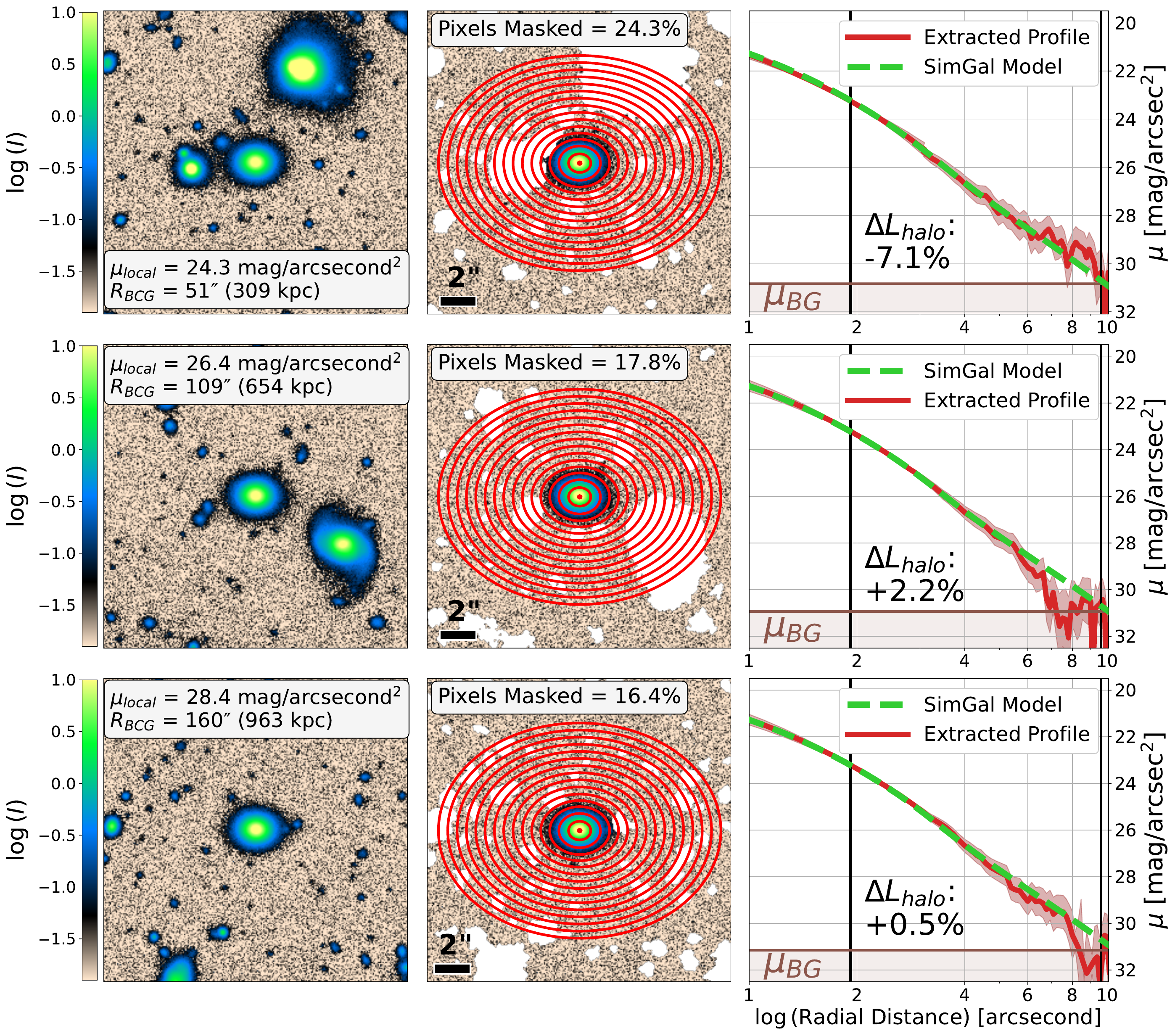}
\caption{\footnotesize{Demonstration of our light profile extraction technique on a simulated galaxy inserted into an HSC-SSP image. Rows show the same galaxy at different cluster locations, with lower rows corresponding to lower-density regions (larger clustercentric distances). Left: $250 \times 250$ pixel cutout of a PSF-convolved simulated galaxy placed in a HSC-SSP $r$-band image. Legends in each panel contain the galaxy's clustercentric distance ($R_{BCG}$) and total local surface brightness ($\mu_{local}$, Section \ref{sec:methods_dense_enviro_tests}). Middle: Final masked and corrected image fitted with isophotes (red ellipses) used to extract azimuthally-averaged $\mu$ values along the direction of the galaxy's semi-major axis (note final profiles are circularized, Section \ref{sec:methods_image_corections_profile_extractions}). The number of isophotes shown has been reduced for visual clarity, but in practice, isophotes are more finely spaced. The percentage of total pixels that are masked (white) is displayed in the legends. Right: Comparison of the initial simulated light profile (green) and the final \texttt{GalPRIME}-extracted light profile (red). The brown horizontal line indicates the average background surface brightness ($\mu_{BG}$) in the final images (middle panels). The black vertical lines indicate $2R_e$ and $10R_e$ for the galaxy and denote our defined stellar halo region. To quantify the performance of the extraction, we show the offset in integrated stellar halo luminosity ($L_{halo}$, Section \ref{sec:methods_stellar_halo}) where $\Delta L_{halo} = [(L_{truth}{-}L_{extracted})/L_{truth}]\cdot 100$.}
\label{fig:isophote}}
\end{figure*}

Following the methodology established in DJW2025, we implement image corrections and extract galaxy light profiles using \texttt{GalPRIME} \citep{new-harrison}, a parallelized Python suite for non-parametric galaxy light profile extractions. In this subsection, we briefly summarize the key aspects of our procedure and demonstrate the overall profile extraction technique using a simulated cluster galaxy (Figure \ref{fig:isophote}).

As in DJW2025, we specifically extract rest-frame $g$-band $\mu$ profiles (or $\mu_g$ profiles) of galaxies over the full redshift range (\zrange). This emission ($\sim 4000{-}5500$ \AA) traces the long-lived lower-mass stars that form the bulk of a galaxy's stellar mass \citep{Spavone2017, Spavone(2021), Huang(2018), Gilhuly(2022)}. We extract individual light profiles from different $grizy$ images depending on a galaxy's redshift ($z$) to trace the same approximate wavelength range (centered at $\sim 5000$\ang), following $\lambda_g$ = $\lambda_{obs}\cdot (1+z)^{-1}$ where $\lambda_g$ and $\lambda_{obs}$ represent rest-frame $g$-band and observed-band wavelengths, respectively. 

As galaxy outskirts are regions of low $\mu$, we must correct for light contamination from foreground and background sources as well as any residual sky-subtracted background noise \citep{Szomoru2012, Trujillo2016, Gilhuly(2022)}. To address this, we implement both the source masking and 2D background subtraction procedures in \texttt{GalPRIME} on individual galaxy images before light profile extraction. Additionally, we apply the forward-modeling point spread function (PSF) correction procedure of DJW2025 to remove the filter-dependent PSF effects that suppress central $\mu$ levels in galaxies and redistribute this light to larger radii \citep{Trujillo-PSF-2001, Dejong2008, Borlaff2017}. All three procedures - source masking, 2D background subtraction, and PSF correction - have been tested and optimized using both simulated data and HSC-SSP images \citep{DevinP1, new-harrison}.

Galaxy light profiles are extracted from the corrected images using \texttt{GalPRIME}'s implementation of the elliptical isophote analysis method of \cite{Jedrzejewski(1987)}. This method, illustrated in Figure \ref{fig:isophote}, fits elliptical isophotes to an image (red circles in the second column of Figure \ref{fig:isophote}) to represent the observed $\mu$ distribution of a galaxy as a function of distance from its center. The fitting procedure allows for variations in the ellipticities and position angles of isophotes with increasing galactocentric distance. Points in the radial light profile (third column in Figure \ref{fig:isophote}) represent azimuthally-averaged $\mu$ values from individual isophotes at those semi-major axis distances. Uncertainties on individual $\mu$ values grow larger with increasing radii, typically ranging from $\sim 0.05$ mag/arcsec$^{2}$ in inner regions ($2R_e$, first vertical black line in Figure \ref{fig:isophote}) to $\sim 1$ mag/arcsec$^{2}$ in the outermost regions of profiles ($10R_e$, second vertical black line in Figure \ref{fig:isophote}).

Extracted profile $\mu$ values (AB magnitudes/arcsecond$^2$) are corrected for cosmological surface brightness dimming ($\mu\propto (1+z)^{-3}$ when using AB magnitudes; \citealt{Whitney2020}). We use absolute solar magnitudes in $grizy$ bands from \cite{sun-filters} to convert $\mu$ values into $L_{\odot}$/pc$^2$ units. Additionally, we convert the major axis profiles that are given by \texttt{GalPRIME} into circularized profiles via $R=\sqrt{ab}$, where $a$ and $b$ represent the major and minor axes, respectively \citep{Graham(2005)}.

\subsection{Testing Profile Extractions in Dense Environments}\label{sec:methods_dense_enviro_tests}
As discussed in Section \ref{sec:methods_image_corections_profile_extractions}, reliably measuring the low $\mu$ levels in galaxy outskirts requires accurate background subtraction and sufficient masking of nearby sources. This is particularly important in galaxy clusters, where the high density of cluster members may complicate source masking. In addition, the diffuse ICL contributes to the observed surface brightness throughout cluster environments. In this analysis, we treat the ICL as a background source that must be removed to isolate the stellar halo light of individual cluster member galaxies.

To test the robustness of our masking and 2D background subtraction procedures (Section \ref{sec:methods_image_corections_profile_extractions}) in dense environments, we perform tests using $100,000$ simulated galaxies consisting of PSF-convolved two-component \sersicS models placed within cluster regions in real HSC-SSP images (one example shown in each row of Figure \ref{fig:isophote}). For each test and simulated cluster galaxy, we run our full \texttt{GalPRIME} light profile extraction pipeline (Section \ref{sec:methods_image_corections_profile_extractions}) and compare the recovered light profile to the initial simulated light profile. We quantify the performance of each test by measuring the fractional offset in integrated stellar halo luminosity ($L_{halo}$, Section \ref{sec:methods_stellar_halo}) between the two profiles, where $\Delta L_{halo} = [(L_{truth}{-}L_{extracted})/L_{truth}]\cdot 100$. In this section, we briefly summarize the key conclusions of these tests and how they affect our final galaxy samples. Appendix \ref{appendixA} includes additional details and figures.

Across the large simulated galaxy sample, we find a very low median offset in $L_{halo}$ ($\Delta L_{halo} \sim 1\%\pm5\%$), indicating that stellar halo luminosities can be reliably recovered in dense environments out to 10$R_e$ (our outermost profile limit, Section \ref{sec:methods_median_profiles}). This demonstrates that \texttt{GalPRIME} adequately removes excess background emission due to the ICL, without explicitly modeling the ICL as a separate component. This demonstrates that \texttt{GalPRIME} adequately removes the diffuse ICL component treated as excess background emission in this analysis, without explicitly modeling the ICL as a separate component.

However, the tests also revealed that in some rare cases a galaxy's extracted light profile and estimated $L_{halo}$ can be significantly offset (e.g., $\Delta L_{halo}\sim 40-50\%$) from the ground truth. We investigate whether any correlations exist between larger $L_{halo}$ offsets and different galaxy properties - total luminosity, $R_e$, and concentration ($C = 5\log(R_{80}/R_{20})$; \citealt{Conselice2003_concentration}) - and environmental parameters (Appendix \ref{sec:appendix_A1}). To quantify environment, we measure the total local surface brightness ($\mu_{local}$) within a circular aperture ($R=125$ pixels) centered on a galaxy's position but with the galaxy's light removed. Additionally, we compute the fraction of total image pixels that are masked and the distance from a galaxy's center to the nearest masked pixel, both indicators of how crowded the surrounding environment is. Simulated galaxies with the largest offsets tend to be fainter and found in denser environments (discussed further in Appendix \ref{sec:appendix_A1}).

To identify any similarly problematic galaxies in our CLAUDS+HSC-SSP sample, for each real galaxy, we find the nearest 10 simulated galaxies in a multi-parameter space (i.e., the six properties described above). From these matched simulated galaxies, we calculate the distance-weighted mean $\Delta L_{halo}$ and assign it to the real galaxy. If any real galaxy has an expected $\Delta L_{halo}$ larger than $1\sigma$ on median $L_{halo}$ values, we remove it from our final galaxy sample due to unreliable $L_{halo}$ measurements. To be consistent across our whole sample, we perform this matching$+$removal procedure on both the field and cluster QG samples. 

The CLAUDS+HSC-SSP sample galaxies with the largest expected $L_{halo}$ offsets due to their similarity with poor-performing simulated galaxies are those that are fainter, smaller in size, and have larger concentrations ($C = 5\log(R_{80}/R_{20})$; \citealt{Conselice2003_concentration}). They also reside in brighter local environments (i.e. they are embedded in an excess of background or ICL with large $\mu_{local}$ values in Figure \ref{fig:isophote}), and have a larger number of masked pixels (second column in Figure \ref{fig:isophote}) and masks closer to a galaxy's center. These are clear indicators of a crowded local environment.

In total, we remove 90 cluster galaxies ($\sim 4.1\%$ of total) and 1985 field galaxies ($\sim 2.1\%$ of total) from our samples due to unreliable light profile measurements. The final sample sizes reported in Section \ref{sec:data_final_sample_selection} (i.e., 2,168 cluster QGs and 94,479 field QGs) and the subsample bin counts in Table \ref{table:MZ_table} already exclude galaxies with unreliable $L_{halo}$ measurements.

\subsection{Median Profiles of Different Galaxy Subsamples}\label{sec:methods_median_profiles}
\begin{figure}[ht]
\centering
\includegraphics[width=0.47\textwidth]{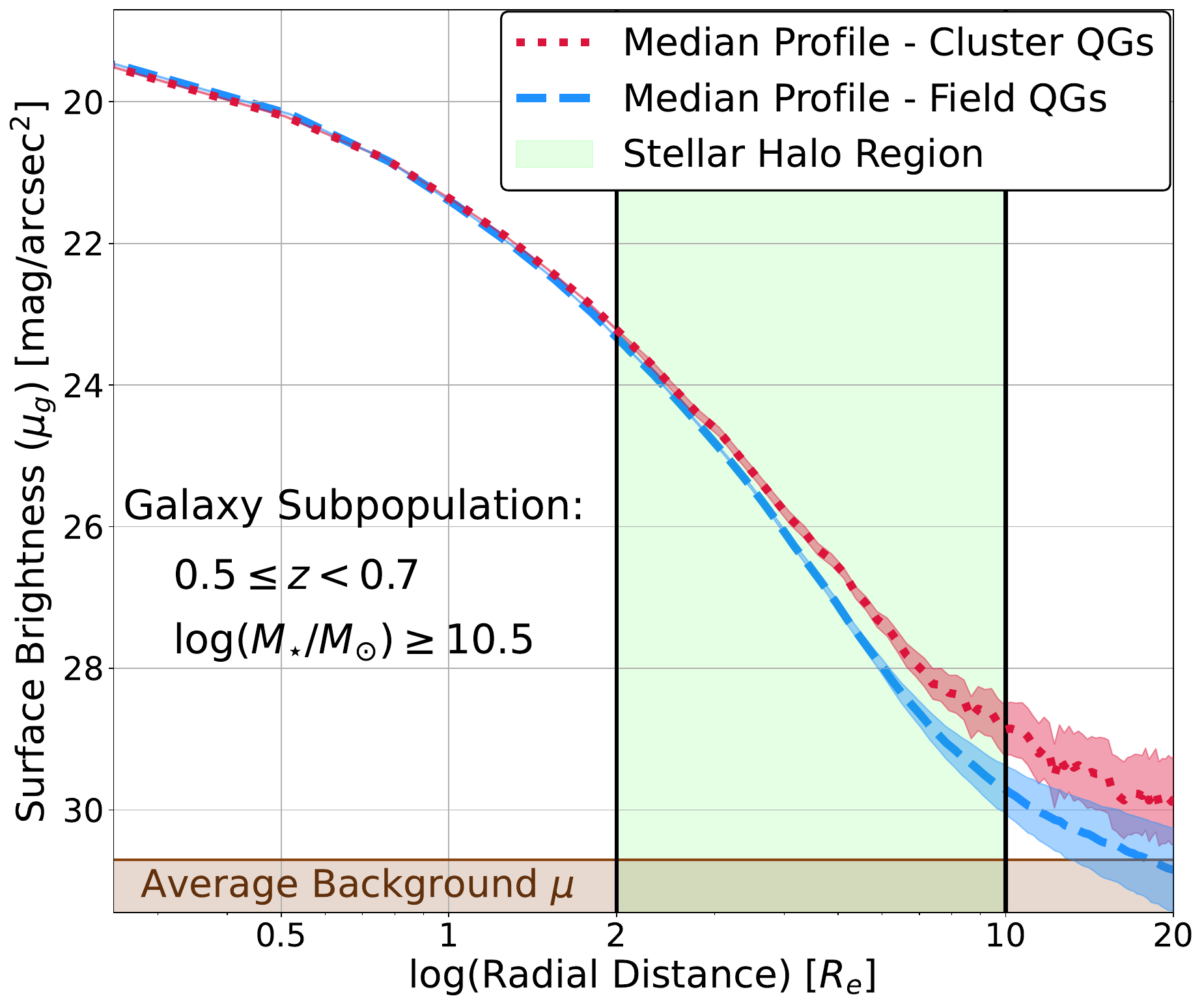}
\caption{\footnotesize{Median $\mu_g$ profiles for field (blue) and cluster (red) QGs of \highmassS at \zthree. Black vertical lines at $2R_e$ and $10R_e$ enclose our defined stellar halo region (shaded in green) and the portion of profiles integrated to calculate $L_{halo}$. The brown shaded region represents the average background $\mu$ level at these redshifts. Errors on median profiles (red and blue shaded regions) are obtained using a Monte-Carlo resampling procedure that propagates individual galaxy measurement errors (Section \ref{sec:methods_median_profiles}). We extend the x-axis beyond $10R_e$ (our radial limit) to demonstrate how errors grow larger with increasing radii.}
\label{fig:median_profile_halo_region}}
\end{figure}

As established in Section \ref{sec:data_final_sample_selection}, we divide our cluster and field QG samples into smaller subsamples (i.e. \msS + $z$ bin combinations, Table \ref{table:MZ_table}) to study stellar halo assembly trends across subpopulations. For each field and cluster QG subsample, we compute median $\mu_g$ profiles. Figure \ref{fig:median_profile_halo_region} compares the median $\mu_g$ profiles of field (blue) and cluster QGs (red) within one of our subsamples (\highmassS and \zthree).

As in DJW2025, we limit each median profile to $10R_e$, which ensures $\mu$ levels in median profile outskirts are at least $\geq3 \sigma$ above the average background level at a given redshift ($\sim 30.2-31.0$ mag/arcsec$^2$), with this offset improving in the lower-mass subsamples (i.e., $\geq 5\sigma$). To estimate uncertainties on the median $\mu_g$ profiles (red and blue shaded regions in Figure \ref{fig:median_profile_halo_region}), we improve on the methodology of DJW2025 by using a new Monte-Carlo resampling procedure that incorporates the $\mu$ measurement errors of individual galaxies. For each galaxy subsample, we generate $N=500$ resampled realizations of the batch of light profiles by drawing galaxies with replacement and perturbing each profile by $1\sigma$. We then recompute the median $\mu_g$ profile for every realization, and the scatter among these resampled medians provides the $1\sigma$ uncertainty on median $\mu_g$ profiles.

Figure \ref{fig:median_profile_halo_region} demonstrates a clear difference in the light contained within a stellar halo region (green shaded region in Figure \ref{fig:median_profile_halo_region}) defined as $2{-}10 R_e$ (Section \ref{sec:methods_stellar_halo}) between field and cluster QGs in this \msS and $z$ range. This difference can be seen as intrinsic and not driven by imperfectly subtracted light contamination in cluster regions (e.g., from ICL), as shown by the simulation-based tests of our profile extraction procedure (Section \ref{sec:methods_dense_enviro_tests} and Appendix \ref{appendixA}). We investigate the differences in stellar halo luminosity between all field and cluster QG subsamples in more detail throughout Section \ref{main-sec:results}.

\subsection{Quantifying Stellar Halo Evolution}\label{sec:methods_stellar_halo}
As in DJW2025, we define the stellar halo region in galaxies to span $2{-}10 R_e$ (green shaded region in Figure \ref{fig:median_profile_halo_region}). This region corresponds to different physical radial ranges depending on the galaxy subsample, ranging from $\sim 4.7{-}23.5$ kpc for low-mass galaxies at $z=0.7{-}1.0$ to $\sim12.7{-}63.5$ kpc for high-mass galaxies at $z=0.1{-}0.35$. This definition is motivated by previous theoretical studies of simulated galaxy stellar haloes (e.g., \citealt{Pillepich2014, Hirschmann(2015), Cook(2016), Merritt2020}) and enables comparison of our results with predictions for stellar halo assembly from cosmological simulations.

To quantify the stellar halo material contained within the stellar halo region of galaxies, we define $L_{halo}$ to be the integrated luminosity within $2{-}10 R_e$ of a given median $\mu_g$ profile using

\begin{equation}
{L_{halo} = \int_{2R_e}^{10R_e} \mu_g(R)2\pi RdR,
\label{eq:L(R)}}
\end{equation}

\noindent where $\mu_g$ represents rest-frame $g$-band surface brightness (in $L_{\odot}$/pc$^2$ units), $R$ represents radial distance (in parsecs), and the $2 \pi RdR$ factor is the surface area element. Galaxy $R_e$ is defined as the radius where the integrated area under the light profile has reached half of the total light (i.e., $L(R_e)$ = $0.5L_{tot}$). The $R_e$ values used in these calculations are median circularized radii for a given \msS and redshift bin combination (Table \ref{table:MZ_table}), based on individual galaxy radii calculated from circularized $\mu_g$ profiles (Section \ref{sec:methods_image_corections_profile_extractions}) via the curve-of-growth procedure.

Table \ref{table:LHALO} provides median \lhaloS measurements (in $L_{\odot}$) and stellar halo fractions (i.e. $L_{halo}/L_{tot}$) for our field and cluster QG subsamples. These stellar halo fractions are consistent ($\lesssim 1{-}2\sigma$) with predicted stellar halo mass fractions (measured as $M_{\star}(R>2R_e) / M_{\star\ tot}$) of simulated galaxies ($\log M_{\star}\geq 10$) in IllustrisTNG \citep{Merritt2020}. Additionally, a small sample of massive ETGs ($\log M_{\star}\geq 10.8$) observed at $z\sim 0.65$ exhibit stellar halo fractions ($\sim 20{-}40\%$, measured $L(R>10$ kpc$) / L_{tot}$; \citealt{Buitrago(2017)}) similar to our high-mass subsamples.

\begin{deluxetable*}{l|cc|cc}
\tablenum{2}
\tablecaption{Median integrated stellar halo luminosities (\lhalo, Eq. \ref{eq:L(R)}) and stellar halo fractions ($L_{halo}/L_{tot}$) for all field and cluster QG subsamples.}
\label{table:LHALO}
\tablehead{
\colhead{Redshift Bins} \vline
    & \multicolumn{2}{c}{Cluster QG $L_{halo}$} \vline
    & \multicolumn{2}{c}{Field QG $L_{halo}$} \\
\colhead{} \vline
    & \colhead{[$\log(L/L_{\odot})$]}
    & \colhead{[$\%$ of $L_{tot}$]} \vline
    & \colhead{[$\log(L/L_{\odot})$]}
    & \colhead{[$\%$ of $L_{tot}$]}}
\startdata
\hline
\multicolumn{5}{c}{\textbf{Low-Mass Galaxies ($9.66 \leq \log M_{\star} < 10.5$)}} \\
\hline
\zone & 9.56 $\pm$ 0.04 & 28.8 $\pm$ 0.7 & 9.61 $\pm$ 0.02 & 26.3 $\pm$ 0.3 \\
\ztwo & 9.37 $\pm$ 0.03 & 22.3 $\pm$ 0.6 & 9.43 $\pm$ 0.02 & 23.3 $\pm$ 0.3 \\
\zthree & 9.27 $\pm$ 0.03 & 16.8 $\pm$ 0.5 & 9.35 $\pm$ 0.02 & 19.1 $\pm$ 0.3 \\
\zfour & 9.06 $\pm$ 0.04 & 14.9 $\pm$ 0.8 & 9.17 $\pm$ 0.02 & 16.1 $\pm$ 0.5\\
\hline
\multicolumn{5}{c}{\textbf{High-Mass Galaxies ($\log M_{\star} \geq 10.5$)}} \\
\hline
\zone & 10.51 $\pm$ 0.04 & 37.8 $\pm$ 1.1 & 10.37 $\pm$ 0.02 & 33.2 $\pm$ 0.5\\
\ztwo & 10.29 $\pm$ 0.03 & 34.1 $\pm$ 0.7 & 10.18 $\pm$ 0.02 & 30.7 $\pm$ 0.3\\
\zthree & 10.11 $\pm$ 0.03 & 28.1 $\pm$ 0.5 & 10.03 $\pm$ 0.02 & 24.9 $\pm$ 0.2 \\
\zfour & 9.75 $\pm$ 0.04 & 21.9 $\pm$ 0.7 & 9.71 $\pm$ 0.02 & 19.7 $\pm$ 0.5 \\
\hline
\enddata
\end{deluxetable*}

Due to the variety of stellar halo region definitions used in the literature (e.g., see \citealt{Merritt2016, Elias2018, Gilhuly(2022)}), in Appendix \ref{appendixC} we investigate how our results change when using a more extended stellar halo definition (e.g., $5{-}10 R_e$). In summary, the observed trends in stellar halo assembly between field and cluster samples presented in this work (Section \ref{main-sec:results}) remain unchanged when we limit the stellar halo region to more extended radial ranges.

\subsection{Matched Field Control Sample}\label{sec:methods_mass_match}

Differences in \lhaloS (Section \ref{sec:methods_stellar_halo}) between cluster and field QG subsamples, highlighted in Table \ref{table:LHALO}, may be strongly affected by the underlying differences in stellar mass (\ms) and redshift ($z$) distribution between cluster and field QGs. Since many galaxy properties (e.g., SFR, colour, metallicity) depend on both \msS and $z$, controlling for these variables is critical for isolating environmental effects (e.g., \citealt{Ellison2009_MZmatch, Guo2009_MZmatch, Bahe2015_MZmatch, Nantais2016_MZmatch, Vaughan2020_MZmatch, Chen2024_MZmatch}).

We conduct a two-sample Cram\'{e}r-von Mises tests (using \texttt{SciPy}'s \texttt{cramervonmises_2samp} package) on the \msS and $z$ distributions of our field and cluster QG samples. The test evaluates whether the two samples are drawn from the same parent distribution, with $p<0.05$ indicating a statistically significant difference and larger values of $T$ corresponding to greater differences between their cumulative distribution functions. The results of this test show there is a statistically significant difference between the \msS ($T=0.54$ and $p=1.2\times10^{-2}$) and $z$ distributions ($T=118.7$ and $p=3.6\times10^{-8}$) of the two samples. Additionally, we perform the same tests on smaller subsamples of the field and cluster QGs (i.e., within narrower $z$ bins than those in Table \ref{table:MZ_table}), and in all cases the tests confirm that the \msS and $z$ distributions of the two samples remain statistically different. We therefore construct an $M_{\star}+z$ matched field control sample to isolate the influence of cluster environments on the light profiles and stellar halo evolution in QGs.

For each cluster galaxy in our sample, we search for field galaxies within $\pm 1\sigma$ of the selected cluster galaxy's stellar mass and redshift, and randomly select 10 of these $M_{\star}+z$ matched galaxies to add to the field control sample. The large size of our field galaxy sample ensures that every cluster galaxy has at least 10 eligible control galaxies. We repeat this procedure 1000 times, drawing a new random set of 10 control galaxies (with replacement across iterations) for each cluster galaxy in each iteration. Across all iterations, the mean stellar mass and redshift offsets between the cluster and field control galaxies are $\Delta \log M_{\star} \sim0.05$ and $\Delta z\sim0.02$, respectively. For all 1000 realizations of an $M_{\star}+z$ matched field control sample, we calculate new median $\mu_g$ profiles (Section \ref{sec:methods_median_profiles}) for all field QG subsamples (i.e., bins in Table \ref{table:MZ_table}). We take the average profile from these 1000 $M_{\star}+z$ matched median profiles as the final profile for a given field QG subsample and recalculate \lhaloS (Section \ref{sec:methods_stellar_halo}).

\begin{figure*}
\centering
\includegraphics[width=0.95\textwidth]{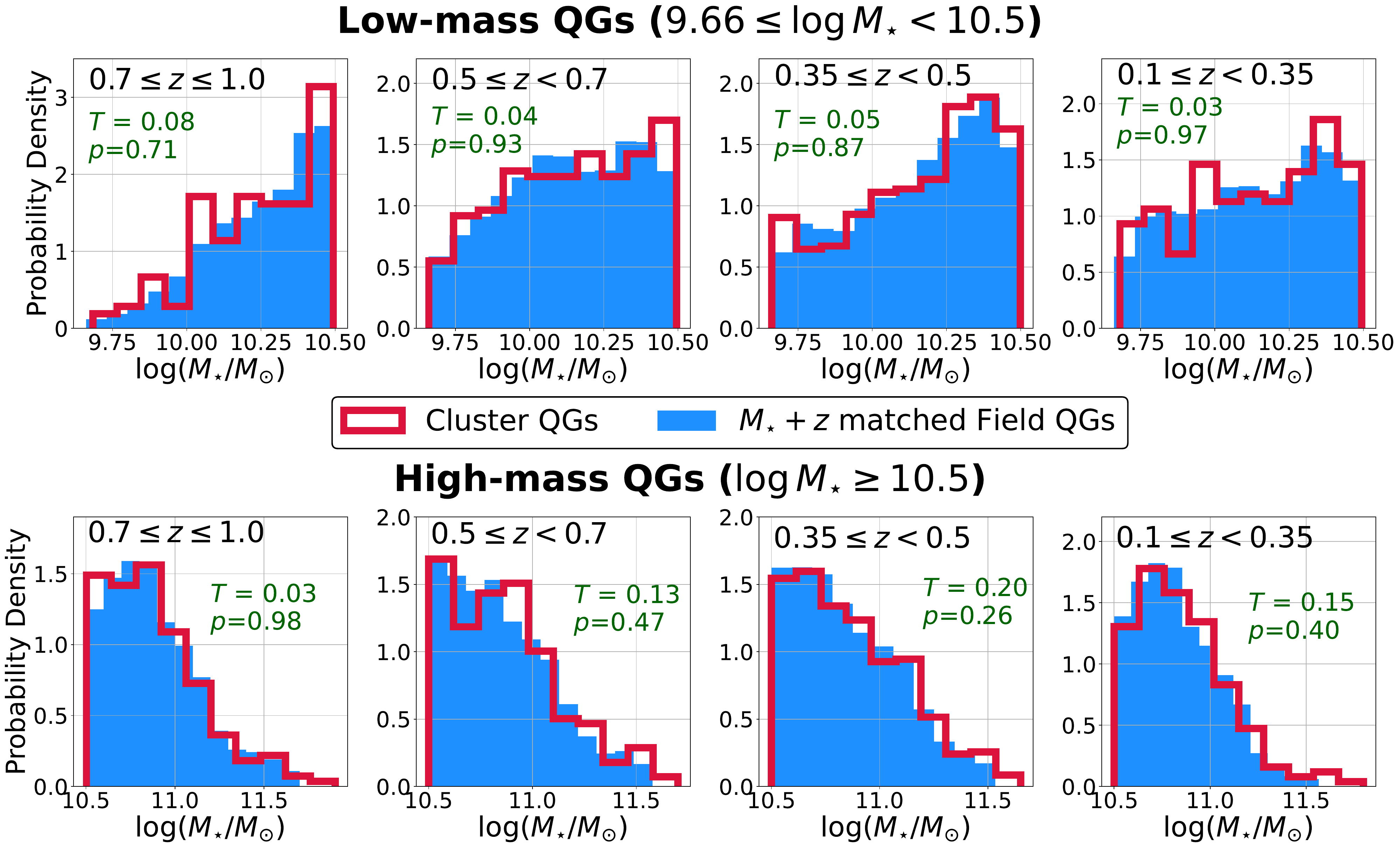}
\caption{Stellar mass distributions of cluster (red) and \ms$+z$ matched field control galaxies (blue) in each of our four redshift bins (separate columns, with ranges given in black text in each panel). The top and bottom rows show our low- and high-mass samples, respectively. The results of two-sample Cram\'{e}r-von Mises tests are given in green text in each panel, demonstrating the distributions are statistically indistinguishable following our \ms$+z$ matching procedure (Section \ref{sec:methods_mass_match}).}
\label{fig:MZ_match_mass_histograms}
\end{figure*}

In summary, we construct an $M_{\star}+z$ matched field control sample to isolate the effect of environment on our results. Figure \ref{fig:MZ_match_mass_histograms} shows the \msS distributions of the cluster (red) and matched field control samples (blue) in each of our final redshift bins (Table \ref{table:MZ_table}). The results of two-sample Cram\'{e}r-von Mises tests (green text in each panel) show that all matched samples have $p>0.05$, indicating that the distributions are now statistically indistinguishable after we perform our matching procedure. Throughout Section \ref{main-sec:results}, we compare the stellar halo evolution of cluster QGs to this field control sample.

\section{Results}\label{main-sec:results}

Here we present the evolution in the integrated stellar halo luminosity (\lhalo, Section \ref{sec:methods_stellar_halo}) of our cluster QG and $M_{\star}+z$ matched field control QG subsamples (Section \ref{sec:methods_mass_match}), derived from median rest-frame $g$-band light profiles ($\mu_g$ profiles, Section \ref{sec:methods_image_corections_profile_extractions} and \ref{sec:methods_median_profiles}).

In Section \ref{sec:results_lhalo_redshift_growth}, we analyze the rate of growth in \lhaloS over \zrangeS in the different subsamples. In Section \ref{sec:results_lhalo_ratio}, we directly compare the median \lhaloS of field and cluster QGs within each redshift interval. Lastly, in Section \ref{sec:results_M200}, we investigate how \lhaloS of cluster QGs depends on the mass of their host cluster ($M_{200}$, Section \ref{sec:data_final_sample_selection}).

\subsection{Rate of Stellar Halo Growth in Field vs. Cluster Galaxy Samples}\label{sec:results_lhalo_redshift_growth}

\begin{figure}[ht]
\centering
\includegraphics[width=0.47\textwidth]{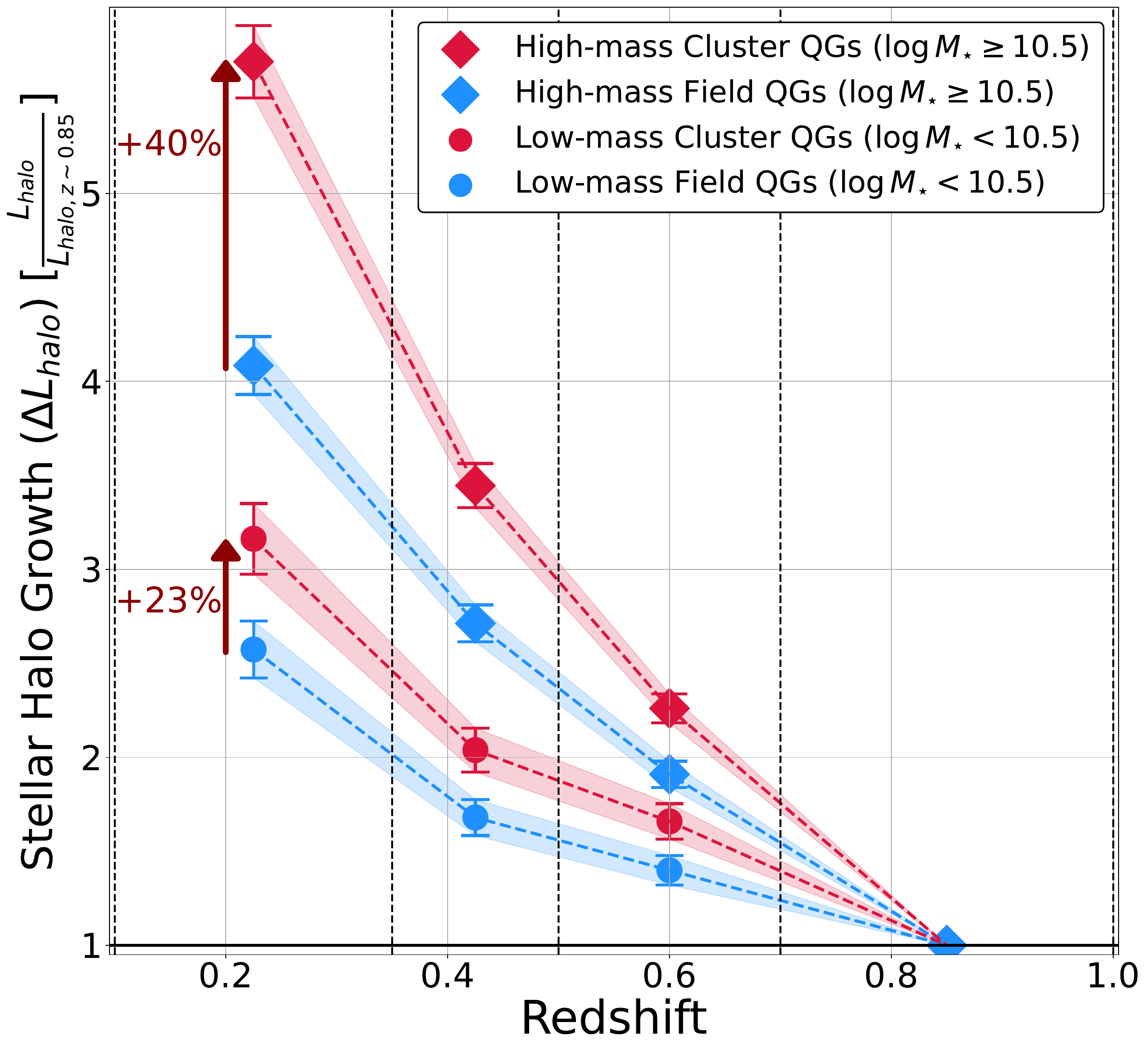}
\caption{\footnotesize Cumulative stellar halo growth (i.e., relative increases in \lhalo) as a function of redshift for field control (blue) and cluster QGs (red). Circles represent low-mass galaxies and diamonds represent high-mass galaxies. Data points represent median \lhaloS values and are placed at the centers of our four redshift bins (Table \ref{table:MZ_table}; and denoted with black vertical dashed lines). Values are normalized to \lhaloS in the highest redshift bin (\zfour; or $L_{halo,\ z\sim 0.85}$). Error bars (and shaded regions) represent Monte-Carlo resampled errors ($1 \sigma$) on median $\mu_g$ profiles (Section \ref{sec:methods_median_profiles}). Cluster QGs grow their stellar haloes more quickly than field QGs over \zrange, and more massive galaxies exhibit greater enhancements in stellar halo growth.
\label{fig:results_halo_growth_factors}}
\end{figure}

In this section, we analyze the buildup of stellar halo material in field and cluster QGs over \zrange. Figure \ref{fig:results_halo_growth_factors} shows how \lhaloS grows with decreasing redshift in the field control (blue) and cluster QGs (red), with values normalized to \lhaloS in the highest redshift bin (\zfour, or $L_{halo,\ z\sim 0.85}$) to more easily compare trends between subsamples. 

High-mass field control QGs (\highmass; blue diamonds in Figure \ref{fig:results_halo_growth_factors}) build up a larger fraction of their stellar halo material than low-mass field QGs (\lowmass; blue circles) over the full $z$-range. The light profile haloes of high-mass and low-mass controls grow by a factor of $4.08 \pm 0.17$ and $2.58 \pm 0.15$, respectively. Although we select the field control sample to match both the \msS and $z$ distributions of our cluster galaxies, we find a good agreement with our previous results for the samples that are representative of the complete field QG population.

As in the field control sample, high-mass cluster QGs (red diamonds in Figure \ref{fig:results_halo_growth_factors}) also build up stellar halo material more rapidly than low-mass cluster QGs (red circles). However, over the same redshift interval, \lhaloS of cluster QGs grows by a factor of $5.70 \pm 0.20$ and $3.16 \pm 0.18$ for the high-mass and the low-mass samples, respectively. Thus, direct comparison of the cluster and field control QG samples reveals that cluster QGs exhibit enhanced stellar halo growth. In the low-mass sample, cluster QGs show a $23\pm 3\%$ larger increase in \lhaloS than field control QGs over \zrange. This positive difference increases to $40 \pm 5\%$ in the high-mass sample.

Previous observational studies of stellar haloes in massive QGs ($\log M_{\star}\gtrsim 11$) report similar trends, where outer stellar mass fractions (e.g., $R>5{-}10$ kpc) grow considerably with decreasing redshift over $0<z<2$ \citep{vanDokkum(2010), Buitrago(2017)}. Our analysis extends across a much wider range in stellar mass, and importantly, reveals that cluster QGs exhibit enhanced stellar halo growth compared to $M_{\star}+z$ matched field galaxies.

The more rapid pace of growth in \lhaloS seen in our cluster sample supports predictions of the environmental effects on halo mass assembly, which suggest low-mass galaxies in higher-density environments exhibit faster mass growth \citep{Christensen2024}. Our findings also align with predictions from the EAGLE simulation \citep{EAGLE}, where cluster member galaxies exhibit mass growth with decreasing redshift and can continue to build up mass for several Gyr after cluster infall (e.g., total \msS can increase by up to $\sim 80\%$ over $\sim 6$ Gyr, \citealt{Sifon2024}).

In summary, we find that cluster QGs exhibit a faster pace of growth in \lhaloS over \zrangeS than $M_{\star}+z$ matched field control QGs. This elevated rate of stellar halo growth is increased in the high-mass cluster sample. In Section \ref{main-sec:discussion}, we discuss the interpretation of our results and the processes that may be driving them.

\subsection{Environmental Differences in Total Stellar Halo Luminosity}\label{sec:results_lhalo_ratio}
In this section, we directly compare median stellar halo luminosities (\lhalo) of cluster and field control QGs within each redshift interval, and investigate how this \lhaloS ratio evolves across \zrange. 

\begin{figure}[ht]
\centering
\includegraphics[width=0.47\textwidth]{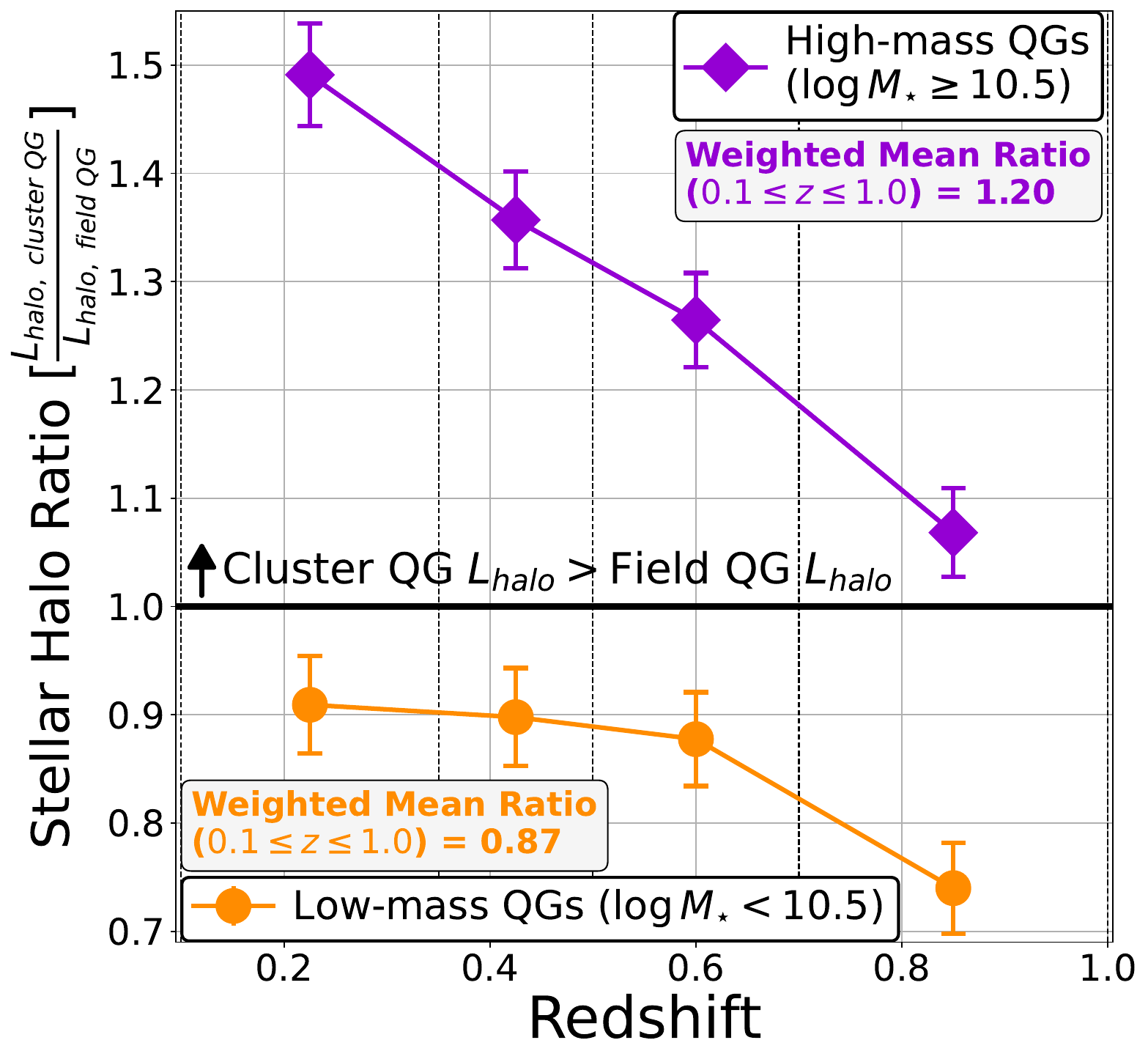}
\caption{\footnotesize{Cluster-to-field \lhaloS ratio as a function of redshift for low- and high-mass galaxies (orange circles and purple diamonds, respectively). Data points represent median values and are placed at the centers of our four $z$-bins (Table \ref{table:MZ_table}; indicated here by vertical dashed lines). The black horizontal line shows the separation between larger \lhaloS in cluster or field QGs. The weighted-mean \lhaloS ratios (Section \ref{sec:results_lhalo_ratio}) across all $z$-bins are given in orange and purple text. Error bars represent propagated Monte-Carlo resampled errors ($1 \sigma$) on median $\mu_g$ profiles (Section \ref{sec:methods_median_profiles}). High-mass cluster QGs host more luminous stellar haloes than the field at a given epoch (redshift range). In contrast, in the low-mass sample, field QGs exhibit larger \lhalo.}
\label{fig:results_halo_ratios}}
\end{figure}

Figure \ref{fig:results_halo_ratios} shows the cluster-to-field \lhaloS ratio (i.e., $L_{halo,\ cluster\ QG}$ / $L_{halo,\ field\ QG}$) as a function of redshift. We weight the \lhaloS ratio by the bin fraction, defined as the number of galaxies in the corresponding redshift bin divided by the total number of galaxies in the sample, and report the $z$-bin-weighted mean \lhaloS ratio in Figure \ref{fig:results_halo_ratios} (orange and purple text for the low- and high-mass samples, respectively).

In the low-mass sample (\lowmass; orange circles), field control QGs exhibit larger \lhaloS in all redshift bins, with a mean \lhaloS ratio of $0.87 \pm 0.04$ across the full $z$-range. The \lhaloS ratio in the low-mass sample increases towards low-$z$, changing from $\sim 0.74$ at \zfourS to $\sim 0.91$ at \zone. 

In contrast to the low-mass sample, high-mass cluster QGs (\highmass; purple diamonds in Figure \ref{fig:results_halo_ratios}) have larger \lhaloS than the field controls with equivalent mass in all redshift bins (mean \lhaloS ratio of $1.20 \pm 0.04$). As in the low-mass sample, the \lhaloS ratio in the high-mass sample also increases toward lower redshifts, but with a stronger trend (e.g., the ratio increases from $\sim 1.07$ at \zfourS to $\sim 1.49$ at \zone.)

Our cluster sample covers the redshift range \zrange. However, several observational studies at low redshift ($z<0.05$) report trends consistent with our results. At these distances, cluster ETGs ($\log M_{\star}\gtrsim 10.5$) exhibit larger $R_{90}$ than QGs in voids, suggesting denser environments result in brighter, more extended outskirts in galaxy light distributions \citep{Perez2025}. \cite{Spavone2020} derived accreted mass fractions for the Fornax Cluster ETGs ($\log M_{\star} \sim 9{-}11$) from multi-component fits to light profile outskirts. Compared to predictions from cosmological simulations (Figure 8 in \citealt{Spavone2020}), these accreted fractions of cluster galaxies lie in the upper half of the predicted distribution across all environments.

Taken together, Figure \ref{fig:results_halo_growth_factors} and Figure \ref{fig:results_halo_ratios} show that while both low- and high-mass cluster QGs grow their stellar haloes faster than field control QGs across \zrange, only high-mass cluster QGs host more luminous stellar haloes at a given epoch. Low-mass cluster QGs start with less luminous stellar haloes than the field controls at \zfourS and their slightly faster growth ($\sim23\%$, Figure \ref{fig:results_halo_growth_factors}) since then does not erase this initial deficit in \lhaloS by \zone. High-mass cluster QGs begin with more luminous stellar haloes and continue to outpace the field, producing an even larger \lhaloS enhancement by \zone.

\subsection{Impact of Cluster Mass on Member Galaxy Stellar Haloes}\label{sec:results_M200}
In this section, we investigate how stellar halo buildup in quiescent cluster galaxies (QCGs) depends on the total mass of their host cluster. More massive clusters host larger galaxy populations and extended infall regions, increasing the likelihood of galaxy-galaxy interactions in their outskirts (e.g., \citealt{Ribeiro2023_cluster_mergers, Kim2024_cluster_mergers, Ivleva2024_cluster_mergers, Edwards2024_cluster_mergers, Watson2025_cluster_mergers}). If these interactions contribute to stellar halo growth (e.g., minor merger-driven accretion), cluster member galaxies residing in more massive host clusters may exhibit enhanced stellar mass buildup. Alternatively, the harsher conditions in more massive clusters may suppress galaxy stellar halo growth. Higher velocity dispersions may hinder mergers and facilitate more high-speed tidal encounters, while denser, hotter ICM components or deeper cluster gravitational potentials may strengthen the effects of disruptive environmental processes (e.g., RPS, tidal stripping; \citealt{Merritt1983, Merritt1985, Moore-1998-harassment, Gnedin2003, Read2006, Boselli2006, Fang2016, Montero2024}).
 
\begin{figure}[ht]
\centering
\includegraphics[width=0.47\textwidth]{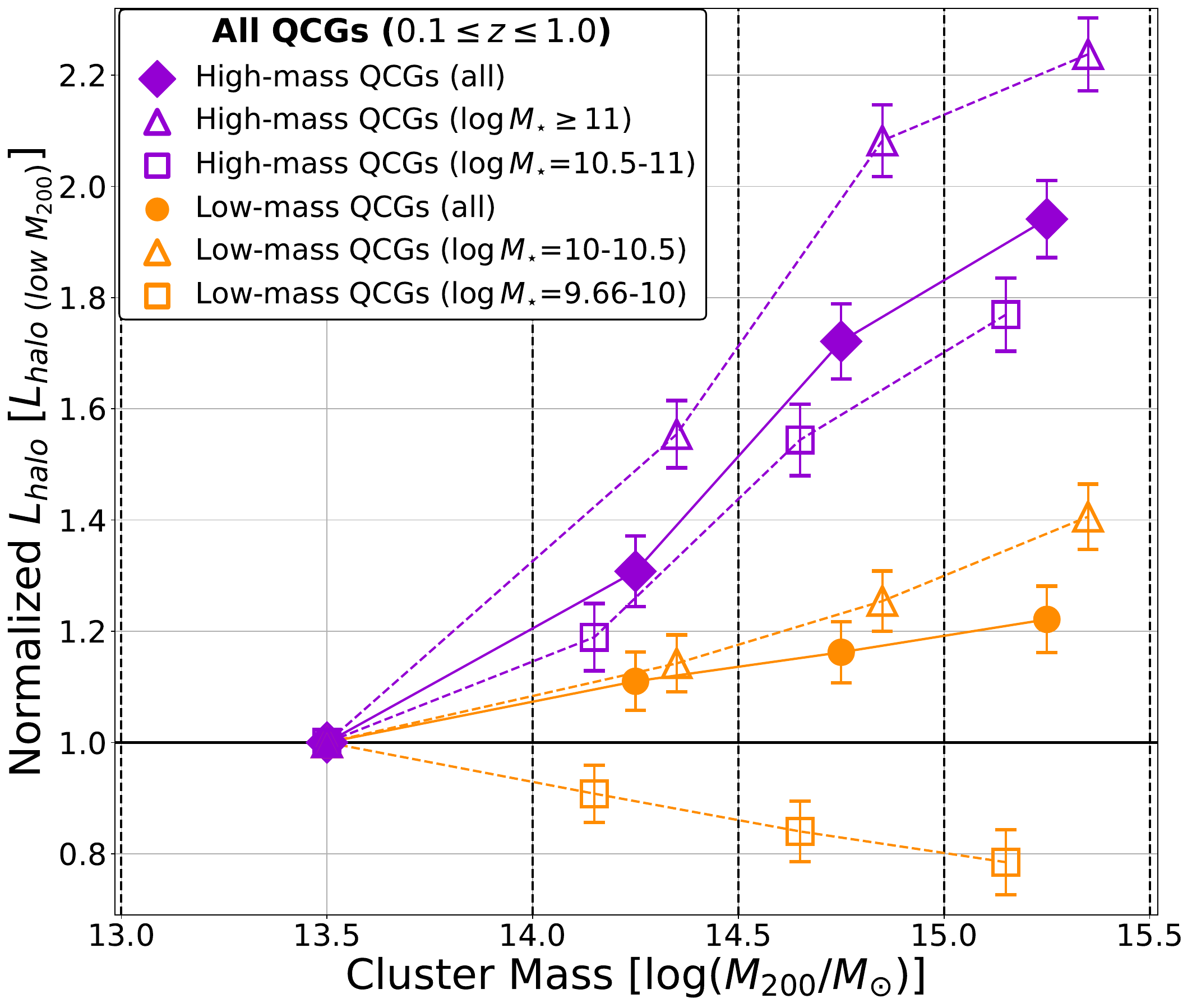}
\caption{\footnotesize{Median \lhaloS as a function of cluster DM halo mass ($M_{200}$, Section \ref{sec:data_final_sample_selection}) for the entire low-mass (solid orange circles) and high-mass (solid purple diamonds) cluster QG samples (i.e., all redshifts combined). Values are normalized to \lhaloS in the lowest $M_{200}$ bin and are placed at the centers of each $M_{200}$ bin (marked by vertical dashed lines). Error bars represent Monte-Carlo resampled errors ($1\sigma$) on median $\mu_g$ profiles (Section \ref{sec:methods_median_profiles}). We split both mass bins into lower-\msS and upper-\msS portions (ranges given in legend), shown as color-coded open squares and open triangles, respectively, and are offset by $\log (M_{200}/M_{\odot}) = \pm0.1$ on the x-axis for visibility purposes. Cluster QGs of $\log M_{\star} \geq 10$ exhibit larger \lhaloS in more massive clusters, with stronger enhancements in more massive galaxies at fixed cluster mass. In contrast, lowest-mass cluster QGs ($\log M_{\star}=9.66{-}10$, orange squares) exhibit smaller \lhaloS with increasing cluster mass.}
\label{fig:results_M200}}
\end{figure}

As described in Section \ref{sec:data_final_sample_selection}, as a proxy for cluster mass, we use DM halo masses ($M_{200}$) obtained from the relation between DM halo mass and BCG stellar mass (Section \ref{sec:data_catalogues}) from \cite{Leauthaud2012_BCG_halomass}. Figure \ref{fig:results_M200} shows the median \lhaloS of low- and high-mass QCGs measured within bins of $M_{200}$ ($\log [M_{200}/M_{\odot}] < 14.0$, $14{-}14.5$, $14.5{-}15$, and $\geq 15.0 $), with values normalized to \lhaloS in the lowest $M_{200}$ bin. Due to the low number statistics, we cannot bin simultaneously by stellar mass, redshift, and $M_{200}$. This analysis combines all galaxies across our four redshift bins (Table \ref{table:MZ_table}). Additionally, we split both the low-mass (\lowmass) and high-mass (\highmass) QCGs into upper-\msS and lower-\msS portions (see legends in Figure \ref{fig:results_M200}) to better understand how results vary across our wide stellar mass range.

High-mass QCGs show a strong dependence of \lhaloS on host cluster mass (solid purple diamonds; Figure \ref{fig:results_M200}), with galaxies in the most massive clusters having $\sim2\times$ more stellar halo material than those in the least massive clusters (\lhaloS larger by a factor of $1.94 \pm 0.07$). This trend is slightly weaker in the lower-\msS portion of the high-mass QCG sample ($\log M_{\star} = 10.5{-}11$, purple squares), with \lhaloS increasing by a factor of $1.77\pm0.06$ between the lowest and highest $M_{200}$ bins. The upper-\msS portion ($\log M_{\star} \geq 11$, purple triangles) instead shows a slightly stronger trend, with \lhaloS increasing by a factor of $2.24\pm0.07$ across the same $M_{200}$ bins.

Low-mass QCGs (solid orange circles, Figure \ref{fig:results_M200}) exhibit a similar positive trend between \lhaloS and cluster mass, albeit much weaker than in the high-mass QCG sample. Low-mass QCGs increase \lhaloS by a factor of $1.22\pm0.06$ between the lowest and highest $M_{200}$ bins. The upper-\msS portion ($\log M_{\star} = 10{-}10.5$, orange triangles) of the low-mass QCG sample exhibits a stronger trend, with \lhaloS increasing by a factor of $1.40\pm0.06$ from low to high $M_{200}$. In contrast, the lower-\msS portion of the low-mass QCG sample ($\log M_{\star} = 9.66{-}10$, orange squares) shows the reverse trend as all other sub-samples, where QCGs found in more massive clusters exhibit smaller stellar haloes. Between the lowest and highest $M_{200}$ bins \lhaloS decreases to a factor of $0.78 \pm 0.06$ (i.e., a $\sim 22\%$ decrease in \lhalo).

Previous observational studies of the environmental dependence of galaxy stellar outskirts have primarily focused on massive central galaxies (e.g., BCGs), generally finding that denser environments or more massive haloes are associated with more prominent galaxy outskirts (e.g., \citealt{Huang2018-DM, Golden-Marx2023, Golden-Marx2025}). We build upon these studies by examining non-central galaxies, finding that the stellar haloes of cluster member QGs (of $\log M_{\star} \gtrsim 10.0$) likewise become more prominent in denser environments. Our results qualitatively agree with low-redshift ($z \sim 0$) studies of cluster ETGs ($\log M_{\star} \sim 9{-}11$), where galaxies with the most extended stellar haloes and largest accreted mass fractions (derived from fits to light profile outskirts) reside in the densest regions of the Fornax cluster \citep{Spavone2020, Spavone2022}.

Our results also support predictions from the Illustris cosmological hydrodynamical simulation \citep{ILLUSTRIS}, where galaxies residing in more massive host dark matter haloes exhibit increased stellar halo luminosity via flatter light profile slopes and larger outer accreted mass fractions \citep{Pillepich2014}.

In conclusion, we find that cluster QGs of $\log M_{\star} \geq 10$ exhibit larger \lhaloS in more massive clusters, with a stronger enhancement seen in more massive galaxies at a fixed host cluster mass. The lowest-mass cluster QGs we study ($\log M_{\star}=9.66{-}10$) instead show a trend of smaller \lhaloS with increasing cluster mass. We discuss the physical interpretation of our results for high-mass and low-mass cluster QGs in Section \ref{sec:discussion_high_mass_interp} and Section \ref{sec:discussion_low_mass_interp}, respectively.

\section{Interpreting Cluster Influence on QG Stellar Halo Assembly}\label{main-sec:discussion}
\subsection{Processes Affecting High-mass QCGs}\label{sec:discussion_high_mass_interp}
Our combined results indicate that high-mass QCGs (\highmass) host more luminous stellar haloes than field control QGs and grow these haloes at a faster pace over \zrange. Additionally, high-mass QCGs found in more massive host clusters (i.e., larger DM halo mass, $M_{200}$; Section \ref{sec:data_final_sample_selection}) exhibit larger \lhaloS than those QCGs found in less massive clusters. We interpret these results as evidence of enhanced stellar halo growth in high-mass QCGs fueled by increased merger-driven accretion in dense environments. 

These mergers are likely minor mergers (or even mini-mergers; e.g., \citealt{Bottrell2024_minimerger, Byrne2025_minimerger, Nipoti2025_minimerger}) as they are more frequent at $z\lesssim1$ (e.g., \citealt{Ownsworth(2014), Rodriguez(2016), Conselice2022}) and mainly deposit material throughout galaxy outskirts where we observe most of the evolution in the median light profiles. Nevertheless, major mergers in cluster environments have been observed (e.g., \citealt{Dokkum1999_cluster_major, Delahaye2017_cluster_major, HyeongHan2025}) and therefore cannot be ruled out. These mergers are also likely predominantly dry, given that our cluster QG sample exhibits lower SFRs and sSFRs, redder colours, and fainter NUV magnitudes (Section \ref{sec:data_final_sample_selection}). More gas-rich (wet) mergers are expected to trigger enhanced star formation and produce bluer colours (e.g., \citealt{Lambas(2012), Hirschmann(2015), Ellison2013, Ellison2018}). To probe the fractional contributions of wet and dry mergers, we will analyze radial colour profiles in future work.

Although mergers are expected to be rare in clusters (particularly in core regions) due to high relative velocities of galaxies (e.g., \citealt{Omori2023, Sureshkumar2024, Yoon2024, angelo_p2_final}), growing observational evidence suggests they can occur in cluster outskirts where conditions are more conducive to lower-speed interactions (e.g., \citealt{Iodice2017, Spavone2020, Ribeiro2023_cluster_mergers, Kim2024_cluster_mergers, Watson2025_cluster_mergers, HyeongHan2025}). Dynamical friction between cluster galaxies can reduce orbital velocities, potentially increasing the likelihood of merging (e.g., \citealt{Goto2005, Ribeiro2023_cluster_mergers}). A related process is dynamical \emph{self}-friction \citep{Miller2020}, in which material stripped from a cluster galaxy can torque its remaining bound material, causing the galaxy to lose orbital angular momentum and become more susceptible to mergers. Using the Three Hundred Project, \cite{Kotecha2021} shows that simulated cluster galaxies in intra-cluster filament regions—where cosmic web filaments connect to clusters—experience reduced environmental effects. The lower ICM gas velocities in these regions enable more efficient accretion of cool gas and may further facilitate mergers.

Galaxies observed in clusters may have assembled some of their stellar halo material through mergers in groups or filament environments before cluster infall (e.g., \citealt{Simha2009, Kim2024_cluster_mergers, Khalid2025_cluster_mergers, Dulcien2026}. Because of this, the observed stellar halo buildup also reflects any pre-infall growth rather than growth arising solely from mergers occurring within clusters after infall. For example, high-mass cluster satellites may previously have been centrals in smaller groups before those groups joined a cluster.

Our interpretation of enhanced merger activity in denser environments is consistent with both observational studies and theoretical predictions of merger fractions across environments. At $0.75<z<1.2$, high-density environments host more dry mergers based on galaxies in the DEEP2 Galaxy Redshift Survey \citep{Lin2010}. At low redshift ($0.075 \leq z \leq 0.2$), a sample of 33,320 galaxies in HSC-SSP data shows enhanced merger activity in denser environments \citep{Yanagawa2025}. Predictions from the Millennium simulation also suggest mergers occur more frequently ($\sim 2.5$ times) in denser regions compared to the least dense regions (e.g., voids) over $0<z<2$ \citep{Fakhouri2009}.

In conclusion, our results support a scenario in which high-mass cluster QGs experience enhanced stellar halo growth over \zrange, driven by increased minor-merger driven accretion in cluster outskirts or in intermediate environments (groups or filaments) before infall. Those high-mass QCGs residing in more massive clusters host more luminous stellar haloes, suggesting they are resilient to disruptive environmental processes (e.g., fly-bys, tidal stripping) and undergo even greater merger-driven growth in denser environments.

\subsection{Processes Affecting Low-mass QCGs}\label{sec:discussion_low_mass_interp}
For the low-mass QCG sample (\lowmass), some of the results are consistent with enhanced stellar halo growth in dense environments fueled by increased merger accretion (as in Section \ref{sec:discussion_high_mass_interp}). This includes the positive trend between \lhaloS and $M_{200}$ for the upper-\msS subsample ($\log M_{\star}=10{-}10.5$; Figure \ref{fig:results_M200}), and the slightly faster pace of growth in \lhaloS of low-mass QCGs compared to low-mass field control QGs over \zrangeS ($\sim 23 \%$, Figure \ref{fig:results_halo_growth_factors}) which is primarily driven by the more massive low-mass QCGs (i.e. $\log M_{\star}\sim 10{-}10.5$). These results support the idea that the most massive galaxies in the low-mass QCG sample (e.g., $\log M_{\star}\sim 10.3-10.5$) likely experience similar assembly histories as the least massive galaxies in the high-mass QCG sample (e.g., $\log M_{\star}\sim 10.6-10.7$).

This increased merger activity in more massive low-mass QCGs is likely occurring in cluster outskirts or in intermediate environments (e.g., group or filament) before cluster infall (as discussed in Section \ref{sec:discussion_high_mass_interp}). This is consistent with predictions from cosmological simulations showing that since $z=1$, low-mass ($\log M_{\star}\sim 9.85{-}10.45$) cluster satellite galaxies experience about half of the total number of mergers before joining a cluster \citep{Simha2009}.

Other results in our low-mass QCG sample, however, indicate that stellar halo growth in the lowest-mass cluster galaxies ($\log M_{\star} < 10$) is inhibited in dense environments. This includes the negative trend between \lhaloS and $M_{200}$ in the lower-\msS subsample ($\log M_{\star}=9.66{-}10$, Figure \ref{fig:results_M200}), and the cluster-to-field \lhaloS ratios of $<1$ (Figure \ref{fig:results_halo_ratios}) which are driven by the lower-mass QCGs in the full low-mass sample. This deficit in \lhaloS seen in lower-mass cluster QGs may be the result of their outer stellar halo stars being stripped through environmental processes and added to the surrounding ICL (e.g., \citealt{Montenegro2023, Contini2018_ICLsim, Contini2024, Brown2024_ICLsim,  Santos2025_ICLsim}). At $z=0.2{-}0.8$, larger amounts of ICL are detected in more massive haloes (e.g., \citealt{Golden-Marx2023, Golden-Marx2025}), consistent with the decrease in \lhaloS with increasing cluster mass for the $\log M_{\star} < 10$ cluster QG subsample (orange squares, Figure \ref{fig:results_M200}). Since light profiles trace stellar distributions, our results support the idea that this is driven by gravitational processes such as high-speed galaxy encounters \citep{Moore1996, Moore-1998-harassment, Bialas2015} and tidal stripping \citep{Toomre1972, Merritt1983, Merritt1985, Read2006, Fang2016}, rather than hydrodynamical processes (e.g., RPS) that affect only the gas within galaxies \citep{Boselli2006, Boselli-RPS}. This interpretation aligns with numerous observational studies which suggest that environmental quenching and stripping become increasingly efficient at lower stellar masses (e.g., \citealt{Peng-2010, Fillingham2016, Fillingham2018, Moutard(2018), Davies2019}).

Additionally, these low-mass cluster QGs can serve as the minor merger companions to more massive cluster galaxies (e.g., \citealt{Lin2010, Yanagawa2025}). In this scenario, the stripped stellar halo material of low-mass QCGs contributes to high-mass QCG mass growth and stellar halo buildup (Section \ref{sec:discussion_high_mass_interp}).

In summary, our combined results support a scenario where more massive galaxies in the low-mass cluster QG population ($\log M_{\star}=10{-}10.5$) experience enhanced stellar halo growth over \zrangeS through increased merger activity occurring in cluster outskirts or in intermediate environments before cluster infall. In contrast, the lowest-mass cluster QGs we study ($\log M_{\star} = 9.66{-}10$) are not growing their stellar haloes inside clusters over \zrange. Instead, they lose some outer stellar halo material to the ICL via environmental stripping or to high-mass galaxies through merger-driven accretion.

\subsection{Additional Processes and Caveats}\label{sec:discussion_additional_processes_caveats}

In addition to our interpretations in Section \ref{sec:discussion_high_mass_interp} and Section \ref{sec:discussion_low_mass_interp}, here we discuss some additional physical processes or observational effects that may be contributing to our measurements, and some caveats to our analysis due to chosen methodology.

An important process linked to galaxy merger activity is merger-induced star formation. During mergers, the tidal forces generated can both compress gas and induce inflows into central regions, which can fuel bursts of both star formation and AGN activity \citep{Ellison2013, Ellison2018, Ellison2020, Wilkinson-merger-starburst, Li2023-starburst}. Cosmological simulations predict that up to $\sim25\%$ of a galaxy's $z=0$ stellar mass can be formed through this merger-induced star formation process (e.g., \citealt{Rodriguez(2016), Husko(2022)}). Based on a sample of 14,000 post-coalescence galaxies at $0.01<z<0.3$, merger-triggered star formation events can lead to a $10-20\%$ increase in a galaxy's stellar mass \citep{Ellison2025_merger_SFR}. However, this additional material resides primarily at smaller radii ($\lesssim 7$ kpc) and therefore may not contribute substantially to the light profile outskirts (and integrated \lhalo) we analyze in this work.

Additional physical processes that may affect galaxy light profile evolution include dynamical heating from external interactions (e.g., \citealt{Hopkins2010-sizemass, Tissera2013, Zhu2022}), internal dynamical processes that drive outward stellar migration \citep{Loebman2011, Loebman2016, Debattista2017}, and adiabatic expansion due to central mass loss from AGN or stellar feedback which can ``puff up" galaxies (e.g., \citealt{Damjanov2009, Trujillo(2011)}). Although these mechanisms can elevate $\mu$ at large radii and mimic accretion-driven growth, their impact on our median $\mu_g$ profiles is expected to be minimal. Based on predictions for galaxy surface mass density profile evolution, adiabatic expansion primarily affects inner galaxy regions and contributes negligibly to profile outskirts (e.g., \citealt{Hopkins2010-sizemass}). Moreover, observational effects - including dynamical heating, inclination differences, and redshift-varying $\mu$ limits - account for a small fraction of the total profile growth over time (e.g., $\leq10\%$ of the mass increase is expected from minor mergers). Additionally, IllustrisTNG predicts that outward stellar migration is partially offset by inward flows, with $\sim23\%$ of the innermost stars in $z=0$ galaxies ($\log M_\star>8.5$) originating at larger radii \citep{Boecker2023}.

Progenitor bias (or newcomer effect; \citealt{Dokkum-1996-PB, Dokkum-2001-PB, Carollo-PB, Damjanov(2019)}) can contribute to the apparent luminosity growth in QG light profiles with decreasing redshift, as recently quenched SFGs—which are larger than QGs at fixed \msS—enter low-$z$ QG samples (e.g., \citealt{Damjanov2022-pivot-PB, Angelo2024, angelo_p2_final}). In DJW2025, we estimated the newcomer effect using fractional changes in QG number densities over $0.2 \leq z \leq 1.1$ derived from stellar mass functions, finding contributions of up to $\sim44\%$ and $\sim12\%$ to the total \lhaloS growth in low-mass ($9.5 \leq \log M_{\star} < 10.5$) and high-mass (\highmass) QGs, respectively. For completeness, we repeat the newcomer effect analysis (Section 6.1.2 in DJW2025) for the new field and cluster QG samples. Owing to the lack of cluster galaxy stellar mass functions in the literature (covering our various redshift intervals), we adopt the same QG stellar mass functions used in DJW2025. We find comparable ($\lesssim2 \sigma$) contributions to $\Delta L_{halo}$ in both environment bins, indicating that the newcomer effect does not drive the environmental differences in stellar halo evolution presented in this work.

A caveat to our analysis is that while we define all luminosity within $2{-}10R_e$ as stellar halo light (Section \ref{sec:methods_stellar_halo}), some fraction of the light at smaller radii within this range may instead arise from inner galaxy components (e.g., \citealt{Trujillo-Chamba, Chamba2022}). This is partially mitigated by our focus on QGs (rather than, e.g., star-forming spirals), although \cite{Gentile2025_euclid} find a higher fraction of QGs ($\log M_{\star} > 9.5$) with stellar disks in high-density environments relative to the field over $0.25 < z < 1$. Thus, a small fraction of the light contributing to \lhaloS in our cluster QG sample could originate from these quiescent disk components. However, our results show only minor quantitative changes when restricting the stellar halo region to larger radii (e.g., $5{-}10R_e$; Figure \ref{fig:discussion_MZmatch_haloregion} in Appendix \ref{appendixC}), where the contribution from disk components would be negligibly small. More generally, our observed trends (Figures \ref{fig:results_halo_growth_factors}-\ref{fig:results_M200}) do not depend on the evolution in galaxy $R_e$. Defining the stellar halo region using a fixed physical radius (e.g., $10 < R  < 50$ kpc, following \citealt{Buitrago(2017)}) yields consistent trends in \lhaloS across cluster and field QG subsamples.

Because our cluster galaxy selection is based on photometric redshifts, there is likely a small fraction of this sample that are redshift interlopers and not true cluster members. However, from tests of our iterative cluster member finding algorithm with the spectroscopically confirmed HectoMAP cluster sample (Section \ref{sec:data_cluster_selection}), we estimate this contamination fraction to be very small (e.g., $\lesssim 4 \%$) based on our final selection of parameter cuts to the cluster member finding code (Section \ref{sec:data_final_sample_selection}). The impact on median $\mu_g$ profiles and our median-based results is negligible from such a small fraction of interloper galaxies compared to the full cluster sample.

\section{Summary and Conclusions}\label{main-sec:conclusions}

In this work, we study how quiescent galaxy (QG) stellar halo assembly is impacted by cluster environments over \zrangeS for a sample of 2,168 cluster and 94,479 field QGs of \MrangeS from the CLAUDS+HSC-SSP photometric catalogs \citep{Desprez2023}. We use deep $grizy$-band images from the Deep/UltraDeep layers of HSC-SSP (PDR3;  \citealt{Aihara(2022)}) to extract galaxy rest-frame $g$-band surface brightness ($\mu_g$) profiles (Section \ref{sec:methods_image_corections_profile_extractions}), enabling us to trace the faint extended stellar halo emission in galaxy outskirts. We study trends in stellar halo assembly by analyzing the evolution in median $\mu_g$ profiles of different galaxy subsamples (i.e., $M_{\star}+z$ bin combinations, Table \ref{table:MZ_table}), and linking the buildup in galaxy light profiles to the underlying stellar mass growth (assuming constant $M_{\star}/L$ ratios throughout galaxies). 

To isolate the effect of environment on our results, we compare the stellar halo evolution in our cluster QG sample with a $M_{\star}+z$ matched field QG control sample (Section \ref{sec:methods_mass_match}). To quantify the effect of clusters on member galaxy stellar haloes, we measure the relative growth in integrated stellar halo luminosity (\lhalo, Section \ref{sec:methods_stellar_halo}) over $0.1<z<1.0$ ($\Delta L_{halo}$). Additionally, we analyze the cluster-to-field \lhaloS ratio ($L_{halo,\ cluster\ QG}/L_{halo,\ field\ QG}$) across various epochs (redshift bins), and how \lhaloS of cluster QGs varies with host cluster mass (i.e., dark matter halo mass, $M_{200}$; Section \ref{sec:data_final_sample_selection}).

We summarize our main results in the following list:

\begin{enumerate}[left=0pt]
    \item In both the cluster and field control samples, high-mass QGs (\highmass) build up stellar halo material faster than low-mass QGs (\lowmass) over \zrangeS (by a factor of $\times$1.6-1.8 over this cosmic time period). Critically, cluster QGs exhibit a more rapid pace of growth in \lhaloS than field control QGs over our full $z$-range (Figure \ref{fig:results_halo_growth_factors}). Low-mass cluster QGs show a $23\pm3\%$ larger increase in \lhaloS relative to their field counterparts over this period. This positive difference increases to $40\pm5\%$ in the high-mass sample. 

    \item High-mass cluster QGs host more luminous stellar haloes (i.e., larger \lhaloS in terms of total $L_{\odot}$) than similarly massive field control QGs at all epochs we observe (purple points in Figure \ref{fig:results_halo_ratios}). The cluster-to-field \lhaloS ratio in the high-mass sample exhibits significant growth with decreasing redshift, increasing from $\sim 1.07$ at \zfourS to $\sim 1.49$ at \zone, with a redshift bin-weighted mean ratio of $\sim1.2$ across the complete $z$-range. In contrast, low-mass cluster QGs host less luminous stellar haloes relative to the field controls in all four redshift bins (orange points in Figure \ref{fig:results_halo_ratios}). The cluster-to-field \lhaloS ratio in the low-mass sample increases from $\sim 0.74$ at \zfourS to $\sim 0.91$ at \zoneS, with a redshift bin-weighted mean ratio of $\sim 0.87$ across the full $z$-range. 

    \item Cluster QGs of $\log M_{\star}\geq 10$ that reside in massive host clusters (with larger $M_{200}$) exhibit larger \lhaloS than similar galaxies residing in less massive clusters (Figure \ref{fig:results_M200}). From the lowest to highest $M_{200}$ bins, \lhaloS is larger by a factor of $\sim1.9$ in high-mass sample (\highmass) and $\sim1.4$ in the upper-\msS low-mass subsample ($\log M_{\star}= 10{-}10.5$).
    
    \item In contrast, the lowest-mass cluster QG subsample ($\log M_{\star}=9.66{-}10$, orange squares in Figure \ref{fig:results_M200}) exhibits the opposite trend: their \lhaloS is reduced in more massive clusters. In these lower-mass cluster QGs, \lhaloS is reduced by $\sim 22\%$ from the lowest to highest $M_{200}$ bins.

\end{enumerate}

Our combined results support a scenario where higher-mass cluster QGs ($\log M_{\star}\geq 10$) undergo enhanced stellar halo growth over \zrangeS relative to the field. This enhanced growth of cluster galaxy haloes is more pronounced at larger \ms. We interpret this buildup of stellar halo material in cluster QGs to be fueled by accretion from increased merger activity in dense environments. This elevated merger growth is likely coming from minor mergers (or mini-mergers, where the mass ratio between the companion and host is $<0.1$; \citealt{Bottrell2024_minimerger}) in cluster outskirts, as cluster cores are expected to hinder merging and cause stronger mass loss in galaxies (e.g., \citealt{Gnedin2003, Montero2024}). Cluster galaxies could have assembled some of their stellar halo material through mergers in groups or filaments prior to cluster infall (e.g., \citealt{Simha2009, Kim2024_cluster_mergers, Khalid2025_cluster_mergers, Dulcien2026}). The observed stellar halo buildup also reflects any pre-infall growth in galaxies, and not only post-infall growth. 
 
In contrast, the stellar halo growth of the lowest-mass cluster QGs we examine ($\log M_{\star}=9.66{-}10$) is hindered in clusters at $z<1$. These lowest-mass cluster QGs appear to be primarily influenced by environmentally-driven stripping of their outer stellar halo material via gravitational processes such as tidal stripping or high-speed galaxy encounters (e.g., \citealt{Moore1996, Moore-1998-harassment, Merritt1983, Merritt1985, Contini2024}). Additionally, some low-mass galaxies may lose stellar material to more massive cluster QGs in minor mergers (e.g., \citealt{Lin2010, Yanagawa2025}). This stellar halo stripping is more efficient in more massive host clusters, likely due to stronger environmental effects caused by deeper cluster potentials and denser ICM components, or from more frequent galaxy-galaxy interactions in the more crowded environments.

In future work, we will extend our light profile analysis to galaxies spanning a continuous range of environmental densities, parameterized using local density estimators. Moving beyond a binary cluster versus field comparison will enable us to assess how stellar halo growth varies across a wider range of environments, potentially revealing where stellar halo growth is most strongly enhanced or suppressed within the cosmic web. We also plan to incorporate an analysis of galaxy rest-frame colour profiles as a function of environment, providing additional insights into the physical processes driving stellar halo evolution. These efforts will be significantly advanced by ongoing and upcoming large spectroscopic surveys of galaxies from instruments such as Euclid \citep{Euclid_mission_intro}, DESI \citep{DESI_1, DESI_2_wen}, Subaru PFS \citep{subaru_PFS}, the Very Large Telescope MOONS \citep{MOONS}, and WHT WEAVE \citep{WEAVE}. Together, these programs will provide both deep imaging and large spectroscopic samples of cluster galaxies extending to higher redshifts than probed here, enabling this analysis to be extended to larger statistical samples and wider ranges of cosmic time. Moreover, their spectroscopic coverage will facilitate robust mapping of galaxy positions within the cosmic web out to $z \sim 2$–3, enabling stellar halo evolution to be studied across a full range of environments, including filaments and voids.

\section*{Acknowledgments}

The research of D.J.W., I.D., and M.S. is supported by the Natural Sciences and Engineering Council (NSERC) of Canada. We utilize computational resources from ACENET and The Digital Research Alliance of Canada. We thank Jubee Sohn for providing valuable insights and suggestions to improve the manuscript.

This work is based on data obtained and processed as part of the CFHT Large Area U-band Deep Survey (CLAUDS), which is a collaboration between astronomers from Canada, France, and China described in \cite{Sawicki(2019)}. CLAUDS data products can be accessed from \url{https://www.clauds.net}. CLAUDS is based on observations obtained with MegaPrime/ MegaCam, a joint project of CFHT and CEA/DAPNIA, at the CFHT which is operated by the National Research Council (NRC) of Canada, the Institut National des Science de l'Univers of the Centre National de la Recherche Scientifique (CNRS) of France, and the University of Hawaii. CLAUDS uses data obtained in part through the Telescope Access Program (TAP), which has been funded by the National Astronomical Observatories, the Chinese Academy of Sciences, and the Special Fund for Astronomy from the Ministry of Finance of China. CLAUDS uses data products from TERAPIX and the Canadian Astronomy Data Centre (CADC) and was carried out using resources from Compute Canada and the Canadian Advanced Network For Astrophysical Research (CANFAR).

This paper is also based on data collected at the Subaru Telescope and retrieved from the HSC data archive system, which is operated by the Subaru Telescope and Astronomy Data Center (ADC) at the National Astronomical Observatory of Japan. Data analysis was in part carried out with the cooperation of the Center for Computational Astrophysics (CfCA), National Astronomical Observatory of Japan. The Hyper Suprime-Cam (HSC) collaboration includes the astronomical communities of Japan and Taiwan, and Princeton University, USA. The Hyper Suprime-Cam (HSC) collaboration includes the astronomical communities of Japan and Taiwan, and Princeton University. The HSC instrumentation and software were developed by the National Astronomical Observatory of Japan (NAOJ), the Kavli Institute for the Physics and Mathematics of the Universe (Kavli IPMU), the University of Tokyo, the High Energy Accelerator Research Organization (KEK), the Academia Sinica Institute for Astronomy and Astrophysics in Taiwan (ASIAA), and Princeton University.
\newline

\textit{Facilities:} Subaru Hyper Suprime-Cam, CFHT MegaPrime / MegaCam  \newline
\textit{Software:} NumPy \citep{numpy}, SciPy \citep{scipy}, Astropy \citep{Astropy}, Photutils \citep{Bradley2022-photutils}, PetroFit \citep{petrofit}, Scikit-learn \citep{scikitimage}, Matplotlib \citep{Hunter-matplotlib}, GalPRIME \citep{new-harrison}

\appendix
\setcounter{figure}{0}
\renewcommand{\thefigure}{\thesection\arabic{figure}}
\section{Testing Profile Extraction Performance in Dense Environments}\label{appendixA}
\subsection{Simulated Galaxy Tests}\label{sec:appendix_A1}

\begin{figure*}
\centering
\includegraphics[width=0.99\textwidth]{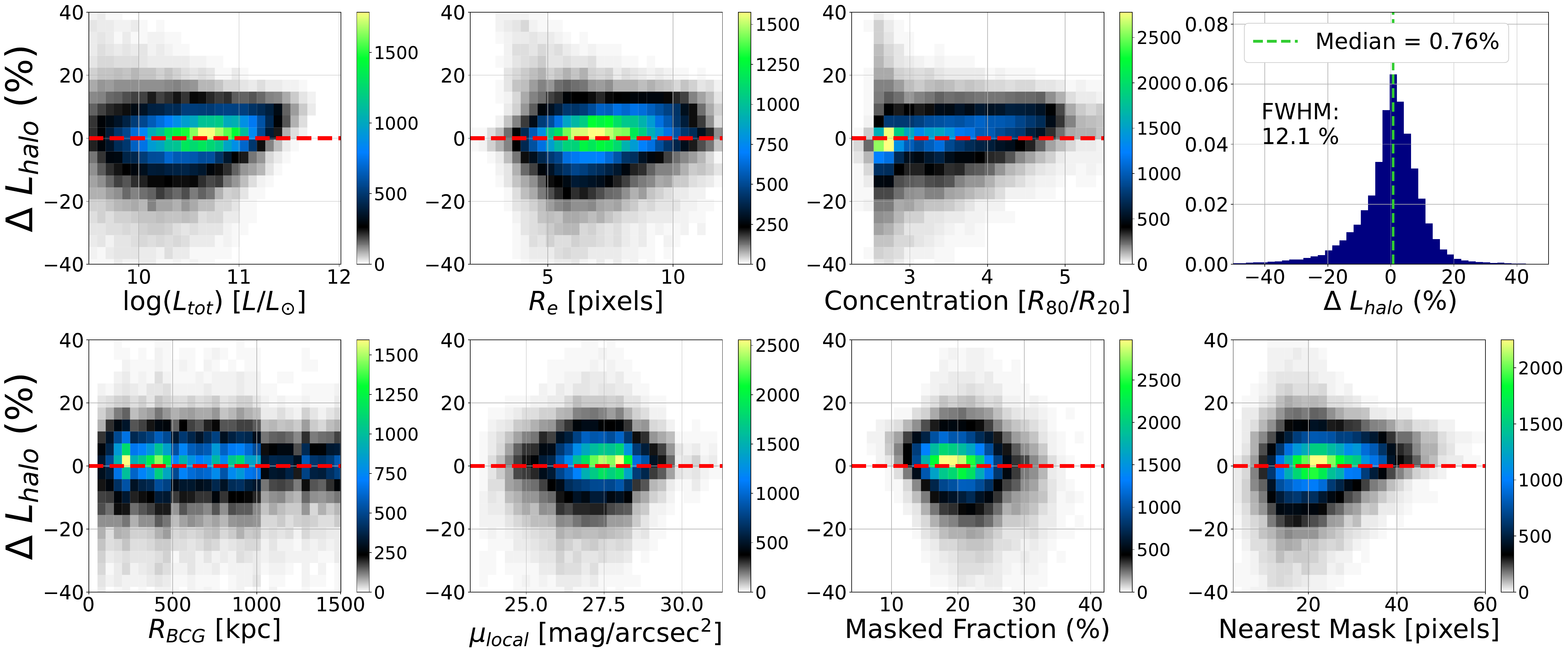}
\caption{\footnotesize{Results of testing our light profile extraction procedure on simulated galaxies placed in cluster environments. Shown in each panel are the integrated stellar halo luminosity offsets ($\Delta L_{halo}$) of individual sim galaxies vs. different galaxy properties or image parameters (described in Section \ref{sec:appendix_A1}). The top right panel shows the total distribution of offsets, as well as the median offset and FWHM.}
\label{fig:appendix_simgal_hist2D}}
\end{figure*}

\begin{figure*}
\centering
\includegraphics[width=0.99\textwidth]{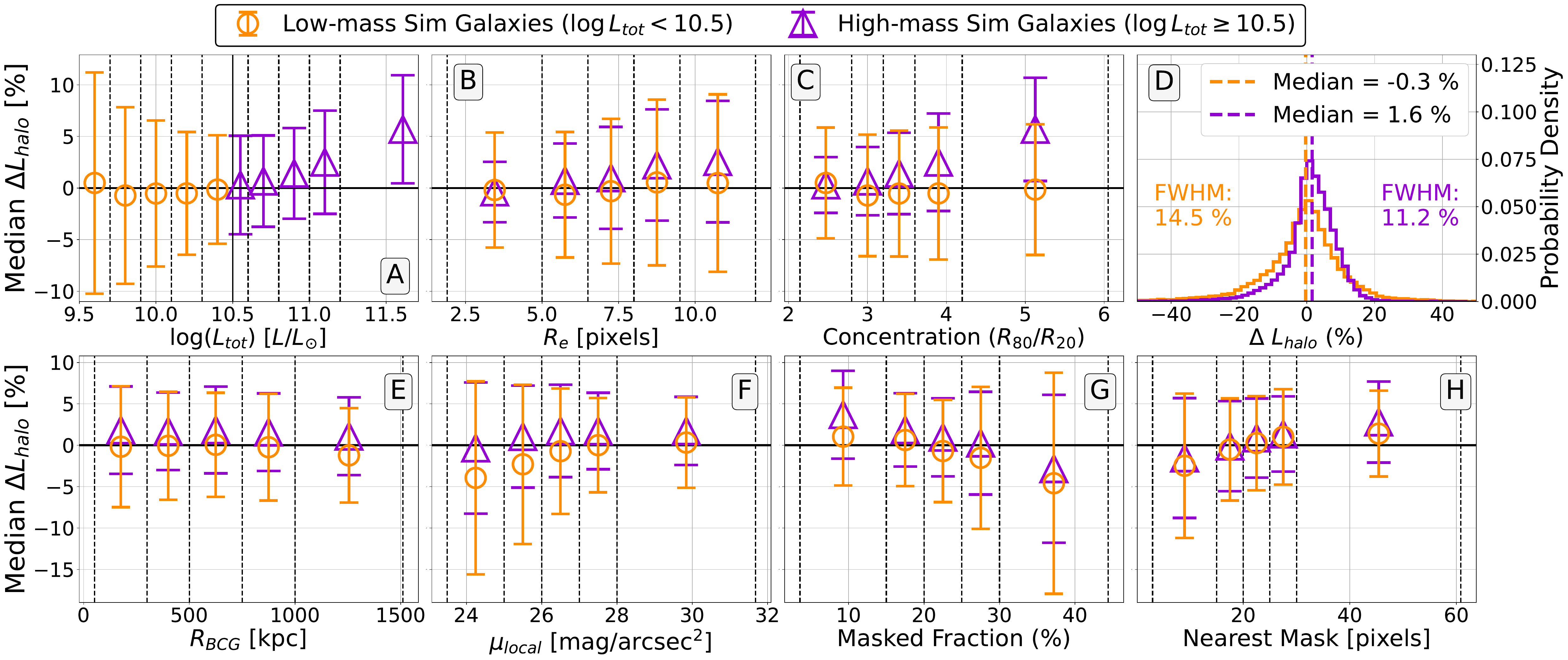}
\caption{\footnotesize{Median $\Delta L_{halo}$ of both low-mass (orange) and high-mass (purple) simulated galaxies measured within smaller bins (indicated by black vertical dashed lines in each panel) of the same properties plotted in Figure \ref{fig:appendix_simgal_hist2D}. Error bars represent $1\sigma$ where $\sigma=(FWHM/2.355)$. Panel D shows the full distribution of individual $\Delta L_{halo}$ split into the two mass bins and provides the median offsets and FWHM for both samples.}
\label{fig:appendix_simgal_binned_offsets}}
\end{figure*}

As discussed in Section \ref{sec:methods_dense_enviro_tests}, we perform a series of tests using simulated galaxies to evaluate the performance of our profile extraction procedure in dense environments. We create 100,000 simulated galaxies consisting of two-component \sersicS models (using \text{astropy}'s \texttt{Sersic2D}), based on the parameters obtained from the 2D decomposition results of \cite{Angelo_P3}, who studied CLAUDS+HSC-SSP galaxies over a similar redshift range. 

Each simulated galaxy is convolved with a HSC-SSP PSF model and randomly placed in an HSC-SSP cutout in the proximity ($25-1500$ kpc) of a real BCG from the sample of \citealt{Oguri2018} (Section \ref{sec:data_cluster_selection}). We then implement our image correction and profile extraction procedure (Section \ref{sec:methods_image_corections_profile_extractions}), and compare the recovered light profile to the initial simulated profile. We quantify the performance of each test by measuring the fractional offset in integrated stellar halo luminosity ($L_{halo}$, Section \ref{sec:methods_stellar_halo}) between the two profiles, where $\Delta L_{halo} = [(L_{truth} {-} L_{extracted})/L_{truth}]\cdot 100$, meaning positive offsets represent an over-subtraction of the true $L_{halo}$. Figure \ref{fig:isophote} (Section \ref{sec:methods_image_corections_profile_extractions}) demonstrates these steps for a single simulated galaxy placed in three different regions of a cluster.

To enable comparison with real CLAUDS+HSC-SSP galaxies (Section \ref{sec:appendix_A2}), we explore the dependence of $\Delta L_{halo}$ on galaxy and image properties measured consistently in both simulations and observations (heatmaps in Figure \ref{fig:appendix_simgal_hist2D}). The galaxy properties we analyze are total luminosity (proxy for \ms), size ($R_e$), and concentration ($C = 5\log(R_{80}/R_{20})$; \citealt{Conselice2003_concentration}). To quantify environment, we measure clustercentric distances ($R_{BCG}$) and the total local surface brightness ($\mu_{local}$) within a circular aperture ($R=125$ pixels) centered on a galaxy's position (with the galaxy's light omitted). Additionally, we compute the distance from a galaxy's center to the nearest masked pixel and the fraction of total image pixels that are masked, both indicative of how crowded an environment is.

Figure \ref{fig:appendix_simgal_binned_offsets} shows median offsets ($\Delta L_{halo}$) measured within bins of the different galaxy and image properties, with error bars representing the width of the distribution of individual offsets in each bin (i.e., $1 \sigma=FWHM/2.355$). Thus, larger error bars signify that extraction performance decreases across the sample (larger individual offsets can occur), regardless of whether the median offsets shift to larger (more negative or positive) values. We find very small median offsets across the large simulated sample, with median $\Delta L_{halo}$ of $-0.3\%$ and $1.6\%$ in low-mass ($\log L_{tot}<10.5$) and high-mass ($\log L_{tot}\geq 10.5$) simulated galaxies, respectively (panel D in Figure \ref{fig:appendix_simgal_binned_offsets}). Despite the near-zero median offsets, however, individual offsets can vary across a fairly large range (e.g., $FWHM\sim 11{-}15\%$). 

Results show fainter objects perform worse, represented by wider distributions of $\Delta L_{halo}$ (larger error bars at lower $L_{tot}$ in panel A in Figure \ref{fig:appendix_simgal_binned_offsets}). We find positive median offsets (i.e., an over-subtraction) for the most luminous simulated galaxies, with a median $\Delta L_{halo}$ of $5\%$ for galaxies of $\log(L_{tot})\geq 11.2$). Larger galaxy sizes also lead to slightly poorer extraction performance (error bars in panel B), but median offsets remain within $-0.5\%$ to $2.5\%$ across all size bins in both low- and high-mass simulated galaxies. No trend between $\Delta L_{halo}$ and galaxy concentration (panel C) is seen in low-mass simulated galaxies, but median offsets in the high-mass sample grow larger with increasing concentration, reaching $\sim6\%$ in the highest concentration bin.

In terms of environment properties, no correlation is found between $\Delta L_{halo}$ and clustercentric distance (panel E, Figure \ref{fig:appendix_simgal_binned_offsets}), with similar median offsets and similar-sized error bars across all $R_{BCG}$ bins. Brighter environments (lower $\mu_{local}$; panel F) lead to worse extraction performance (larger error bars) in both mass bins, as well as a shift towards larger, more negative median offsets (i.e, an under-subtraction) in the low-mass simulated galaxies only. Larger fractions of masked pixels result in poorer extraction performance (wider $\Delta L_{halo}$ distributions) and more negative median offsets in both mass bins (panel G). Additionally, source masks closer to the target galaxy's center also cause a drop in performance and generally lead to more negative median offsets (panel H).

In summary, across a large simulated galaxy sample, our profile extraction procedure performs very well in cluster regions. The resulting impact on our main measured quantity ($L_{halo}$) due to background subtraction and source masking difficulties in dense environments is negligible (median $\Delta L_{halo} \sim 0.8\%$).

\subsection{Refining Observed Sample}\label{sec:appendix_A2}
The results of our simulation-based tests (Section \ref{sec:appendix_A1}) show that in rare cases a galaxy’s extracted light profile and integrated \lhaloS can be significantly offset from true values (e.g., $\Delta L_{halo} \sim 40{-}50\%$). To mitigate the impact of these failures on our CLAUDS+HSC-SSP sample, we matched each real galaxy to its nearest 10 simulated counterparts in a 6D parameter space ($L_{tot}$, $R_e$, $C$, $\mu_{local}$, fraction of masked pixels, and distance to the nearest masked pixel; Appendix \ref{sec:appendix_A1}). From this simulated subsample, we compute a weighted (by $1/$distance) mean $\Delta L_{halo}$ offset. If this offset exceeds the typical $1\sigma$ uncertainty on median $L_{halo}$ values, the galaxy is removed from our final sample due to an unreliable $L_{halo}$ measurement. To be consistent across our whole galaxy sample, we perform this matching and removal procedure on both our field and cluster samples.

\begin{figure}[ht]
\centering
\includegraphics[width=0.47\textwidth]{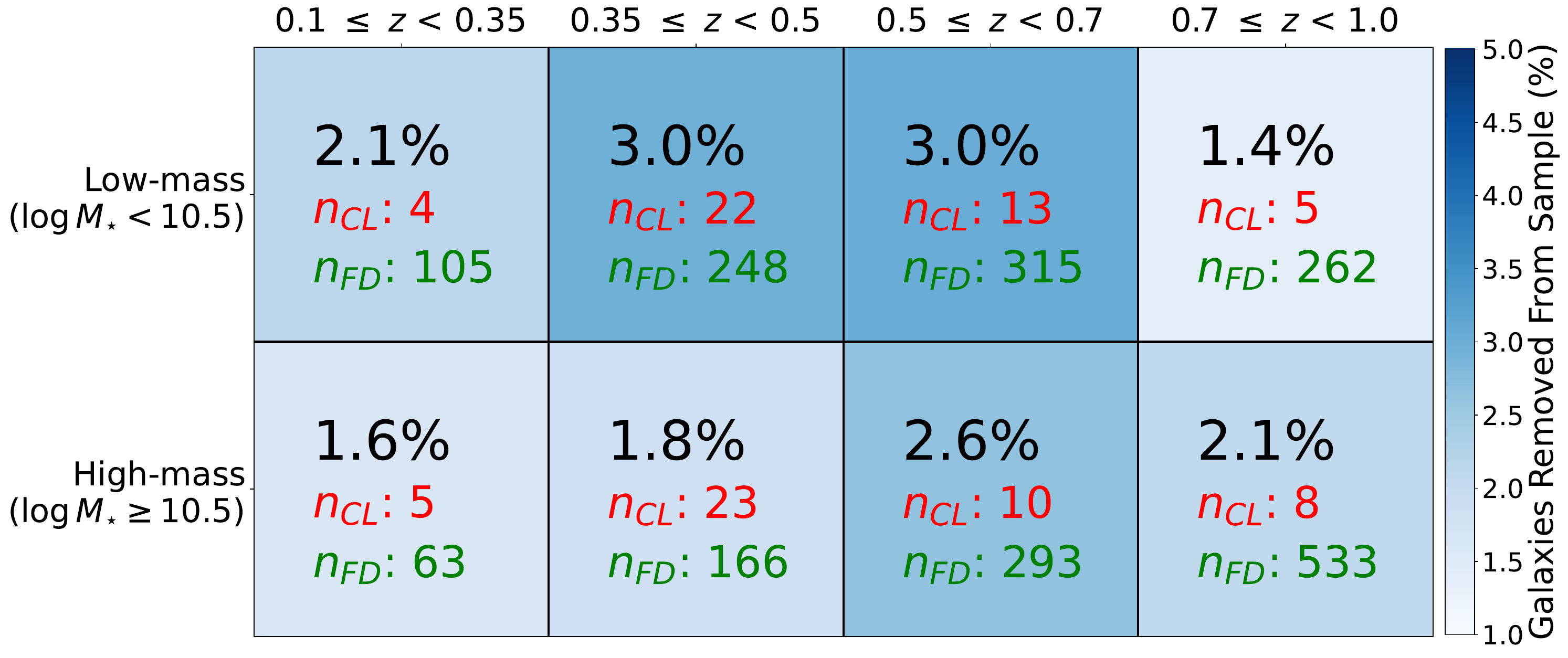}
\caption{\footnotesize{Percentage of galaxies removed from each of our stellar mass and redshift bin combinations due to unreliable $L_{halo}$ measurements stemming from their similarity to poor-performing simulated galaxies. In each panel, we show the percentage removed from the full combined sample (field + cluster), with the specific number of removed cluster ($n_{CL}$) and field ($n_{FD}$) galaxies shown in red and green text, respectively.}
\label{fig:appendix_removal_boxplot}}
\end{figure}

Figure \ref{fig:appendix_removal_boxplot} shows the percentage of galaxies removed from the full combined sample (i.e. cluster + field) in each of our \msS and $z$ bin combinations (Table \ref{table:MZ_table}), along with the specific number of removed cluster ($n_{CL}$) and field ($n_{FD}$) galaxies in each subsample. In total, we remove 90 cluster galaxies ($\sim 4.1\%$ of total) and 1985 field galaxies ($\sim 2.1\%$ of total).

\begin{figure}[ht]
\centering
\includegraphics[width=0.47\textwidth]{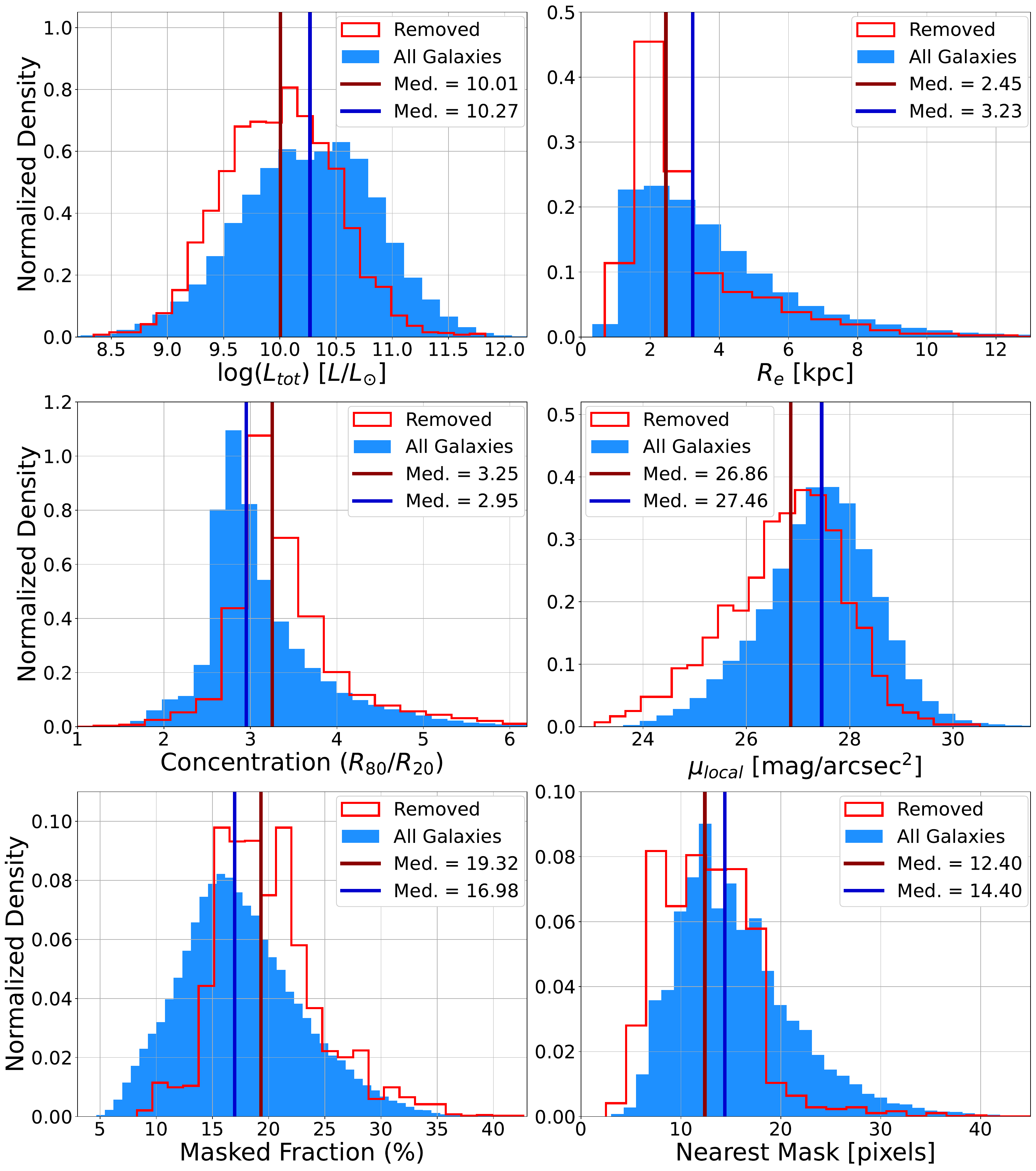}
\caption{\footnotesize{Comparisons of the distributions of the six parameters used in the galaxy removal process (Section \ref{sec:appendix_A2}) between the full galaxy sample (blue histograms) and the galaxies removed due to high expected $\Delta L_{halo}$ (red histograms). Median parameter values are represented by the vertical red and blue lines in each panel, with values reported in the legends.}
\label{fig:appendix_removal_histograms}}
\end{figure}

In Figure \ref{fig:appendix_removal_histograms}, we compare the distributions of the six parameters used in this removal process between the full galaxy sample (blue histograms) and the removed galaxies (red histograms). In conclusion, the galaxies in our CLAUDS+HSC-SSP sample that have the largest expected $\Delta L_{halo}$ based on their similarity with poor-performing simulated galaxies are those that are fainter (change in median $\log L_{tot}$ of $\sim -0.3$ dex), smaller in size (change in median $R_e$ of  $\sim -24\%$), and have larger concentrations (change in median $C$ of $\sim +10\%$). Additionally, they reside in brighter local environments (median $\mu_{local}$ brighter by $\sim -0.6$ mag/arcsec$^2$), have larger fractions of pixels masked (median of $\sim 19.3\%$ vs. $\sim17.0\%$) and have masks which are closer to a galaxy's center (median of $12.4$ vs. $14.4$ pixels, change of $-14\%$).

\section{Effects From Using Global vs. Local Sky-Subtracted HSC-SSP Images}\label{appendixB}
\setcounter{figure}{0}
\begin{figure*}[ht]
\centering
\includegraphics[width=0.99\textwidth]{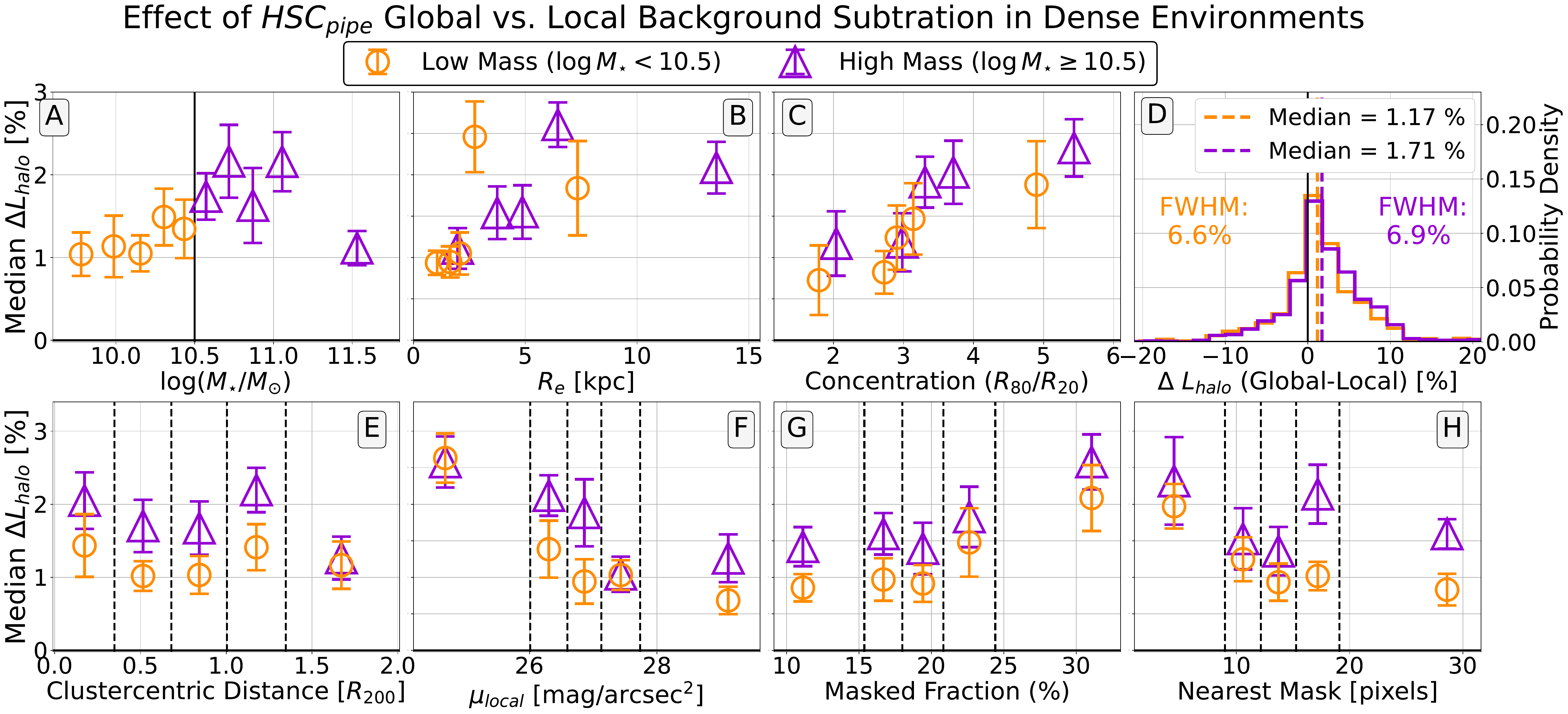}
\caption{Effect on \lhaloS measurements of our cluster sample when using the global- or local-sky subtracted HSC-SSP images. Panel D shows the full distribution of $\Delta L_{halo}$ for the low-mass (orange) and high-mass (purple) samples. Remaining panels show median $\Delta L_{halo}$ measured within bins of different parameters: stellar mass (panel A), galaxy $R_e$ (panel B), galaxy concentration (panel C), clustercentric distance (panel E), local surface brightness ($\mu_{local}$, panel F), fraction of masked pixels in the galaxy's image (panel G), and the nearest masked pixel to the galaxy's center (panel H). Error bars represent bootstrapped uncertainties ($1 \sigma$) on median offsets. Bin widths for the bottom row are the same for the low-mass and high-mass samples (indicated by black vertical dashed lines), while the top row uses equal-size binning unique to each mass bin as they span different ranges of those parameters.}
\label{fig:appendix_global_local_binned_offsets}
\end{figure*}
\begin{figure}[ht]
\centering
\includegraphics[width=0.42\textwidth]{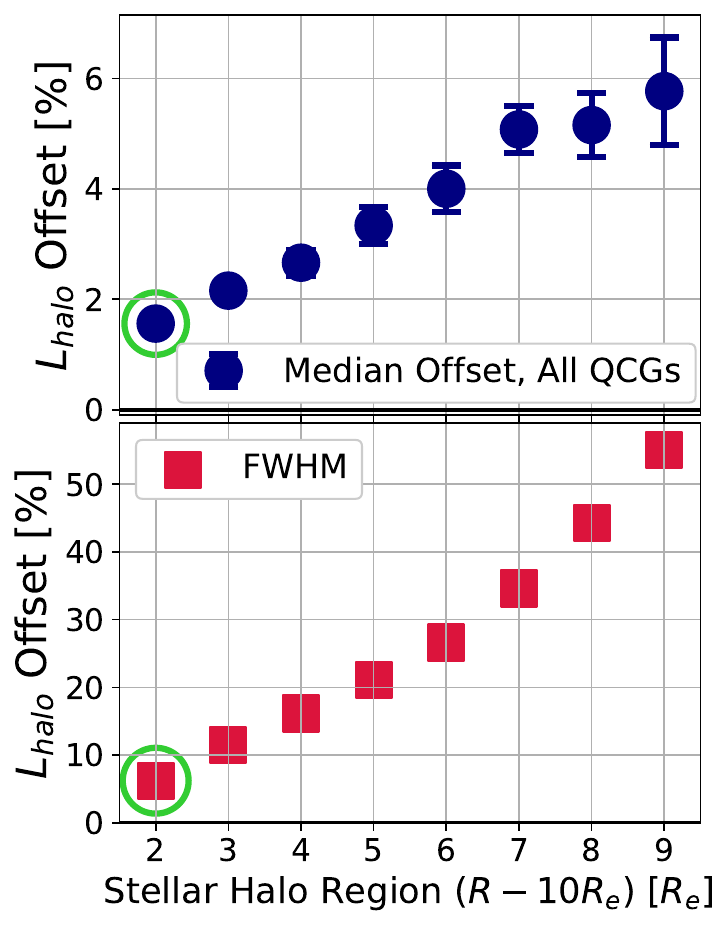}
\caption{Top: Median $\Delta L_{halo}$ between the global- and local-sky subtracted profiles of the full cluster QG sample when the stellar halo region is increased to larger radii (in $R_e$) when integrating galaxy light profiles (Section \ref{sec:methods_stellar_halo}). Bottom: FWHM of the $\Delta L_{halo}$ distribution over the same stellar halo regions. Green circles highlight the results using our original stellar halo region of $2{-}10R_e$.}
\label{fig:appendix_global_local_haloregion}
\end{figure}

As mentioned in Section \ref{sec:data_catalogues}, in this work, we use HSC-SSP images produced with the global-sky subtraction applied rather than the local-sky subtraction procedure. The local-sky subtraction was found to over-subtract the sky around extended objects, and the global-sky subtraction was introduced to correct this effect \citep{Bosch(2018), Aihara2019-PDR2, Aihara(2022)}. Given that we study faint extended stellar halo emission via galaxy light profiles, the global-sky subtracted images are best suited for our analysis. However, for thoroughness, here we demonstrate the effect on the light profile extractions of our cluster sample when using local-sky subtracted images.

For each cluster QG in our sample, we extract their $\mu_g$ profile using a local-sky subtracted HSC-SSP $grizy$ image (the $\mu_{g,\ local}$ profile) following the same procedure outlined in Section \ref{sec:methods_image_corections_profile_extractions}. We then compare the \lhaloS calculated using this $\mu_{g,\ local}$ profile to the \lhaloS calculated using the galaxy's original global-sky subtracted light profile (the $\mu_{g,\ global}$ profile), defining the offset as $\Delta L_{halo} = [(L_{global} {-} L_{local})/L_{global}]\cdot 100$. In Figure \ref{fig:appendix_global_local_binned_offsets} we show the results of these comparisons between global and local profile extractions. Panel D shows the distribution of $\Delta L_{halo}$ values for low-mass (orange) and high-mass (purple) galaxies, with other panels showing median $\Delta L_{halo}$ calculated within bins of the same parameters from Figure \ref{fig:appendix_simgal_binned_offsets} (see Appendix \ref{sec:appendix_A1} for definitions). 

We find very small median offsets of $\Delta L_{halo} = $ $1.17\%$ (FWHM = $6.6\%$) and $1.71\%$ (FWHM = $6.9\%$) in the low-mass and high-mass cluster samples, respectively. No trend is found between $\Delta L_{halo}$ and stellar mass (panel A, Figure \ref{fig:appendix_global_local_binned_offsets}) or clustercentric distance (panel E). Galaxies with larger sizes (panel B) and larger concentrations (panel C) in both mass bins have slightly larger offsets, with median $\Delta L_{halo}$ increasing from $\sim1\%$ to $\sim2.5\%$ across the bins of both parameters. Galaxies in denser (brighter) local environments also exhibit larger offsets, with median $\Delta L_{halo}$ increasing from $\sim1\%$ to $\sim3\%$ across the bins of $\mu_{local}$ (panel F). Similarly, median $\Delta L_{halo}$ increases from $\sim1\%$ to $\sim2.5\%$ across the bins of both masked fraction (panel G) and nearest masked pixel (panel H), where more masked objects or masks closer to a galaxy's center result in larger offsets.

The results in Figure \ref{fig:appendix_global_local_binned_offsets} demonstrate that any impact on our primary results (i.e. \lhaloS measurements) from using global- rather than local-sky-subtracted images is negligible. To better contextualize this analysis for future studies by others, we also examine how $\Delta L_{halo}$ behaves when the stellar halo region (i.e., $2{-}10 R_e$) is pushed to increasingly larger radii when integrating galaxy light profiles. Figure \ref{fig:appendix_global_local_haloregion} shows how the median offset (top panel) and the FWHM of the offset distribution (bottom panel) change as the inner radial boundary is moved outward to larger values of $R_e$. Although median offsets remain small even at very extended radii (e.g., $\sim 6\%$ within 9-10$R_e$), the FWHM grows significantly with increasing radius (e.g., $\sim55\%$ within 9-10$R_e$). This indicates that analyses targeting extreme outskirts are far more sensitive to whether the global- or local-sky subtraction is used, and such effects should be explicitly tested rather than assumed to be negligible.

\section{Effect of Stellar Halo Region Definition}\label{appendixC}
\setcounter{figure}{0}
\begin{figure}[h]
\centering
\includegraphics[width=0.47\textwidth]{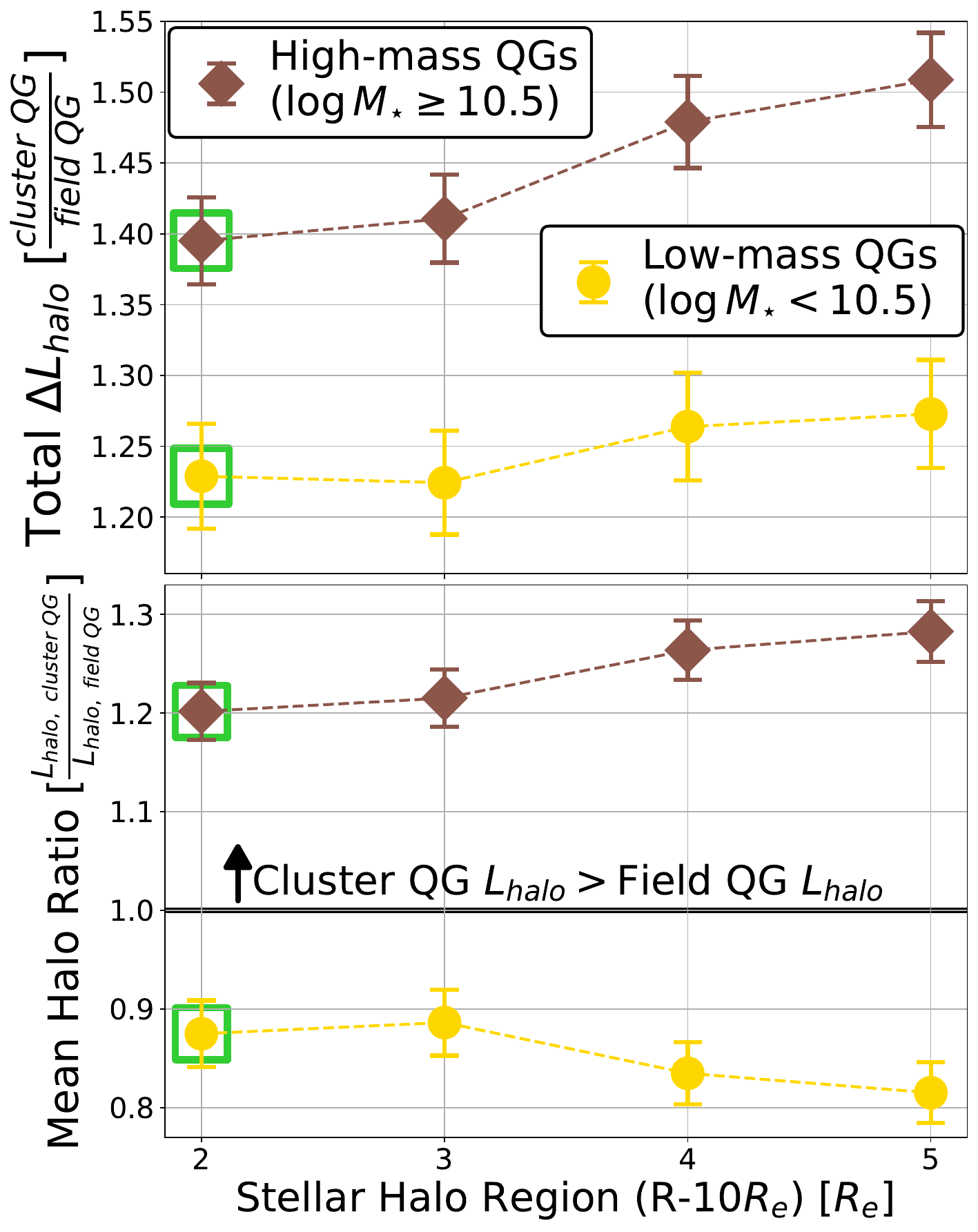}
\caption{Figure shows the effect on two main results (Section \ref{sec:results_lhalo_redshift_growth} and Section \ref{sec:results_lhalo_ratio}) from varying the stellar halo region over which \lhaloS is measured. Top: Comparison of total stellar halo growth over \zrangeS ($\Delta L_{halo}$) in cluster and field QGs (i.e., red percentages in Figure \ref{fig:results_halo_growth_factors}) as a function of different halo region definitions (different $R_e$ ranges, x-axis). Brown diamonds and gold circles represent the high- and low-mass samples, respectively. Points inside green squares represent results using our original stellar halo region of 2-10$R_e$. Bottom: Same design as top panel, but y-axis now shows the bin-weighted mean cluster-to-field \lhaloS ratio ($L_{halo,\ cluster\ QG}$ / $L_{halo,\ field\ QG}$; Figure \ref{fig:results_halo_ratios} in Section \ref{sec:results_lhalo_ratio}). Error bars on all points represent Monte-Carlo resampled errors ($1\sigma$) on median $\mu_g$ profiles.}
\label{fig:discussion_MZmatch_haloregion}
\end{figure}

Throughout this study, we define the stellar halo region of galaxies as $2{-}10 R_e$ (Section \ref{sec:methods_stellar_halo}), motivated by past theoretical studies and to maintain consistency with our previous work (DJW2025). This choice is somewhat arbitrary, however, and a range of definitions have been adopted in the literature (e.g., \citealt{Merritt2016, Elias2018, Gilhuly(2022)}). To understand how our definition impacts our observed trends, throughout this section, we examine how results vary when using a more extended stellar halo region ($3{-}10R_e$, $4{-}10R_e$, and $5{-}10R_e$). 

Figure \ref{fig:discussion_MZmatch_haloregion} shows how varying the stellar halo region affects two of our main results from Section \ref{main-sec:results}: the difference in total stellar halo growth over \zrangeS ($\Delta L_{halo}$) between cluster and field QGs (i.e., red percentages in Figure \ref{fig:results_halo_growth_factors}), and the redshift bin-weighted mean cluster-to-field \lhaloS ratios (i.e., colored text in Figure \ref{fig:results_halo_ratios}). 

In the high-mass sample, restricting the stellar halo region to larger radii leads to larger enhancements in the rate of stellar halo growth between high-mass cluster and field control QGs (brown diamonds, top panel in Figure \ref{fig:discussion_MZmatch_haloregion}), with $\Delta L_{halo}$ increasing by a factor of $\sim1.40$ using $2{-}10R_e$ compared to $\sim1.51$ using $5{-}10R_e$. There is only a minor increase in the $\Delta L_{halo}$ enhancement in the low-mass sample (gold circles, top panel), which could be considered as no change within uncertainties. Our results align with predictions from \cite{Tau2025}, who found only minor changes to the stellar halo accreted mass fractions of simulated galaxies in the Auriga project when using $>4R_e$ or $>5R_e$ as the stellar halo region.

The mean cluster-to-field \lhaloS ratio in the high-mass sample (brown diamonds, bottom panel in Figure \ref{fig:discussion_MZmatch_haloregion}) exhibits minor increases when restricting the stellar halo region to more extended ranges, changing from $\sim1.20$ at $2{-}10R_e$ to $\sim1.28$ at $5{-}10R_e$. The reverse is true in the low-mass sample (gold circles, bottom panel), with the mean \lhaloS ratio changing from $\sim0.87$ at $2{-}10R_e$ to $\sim0.81$ at $5{-}10R_e$. This indicates that in the low-mass sample, probing deeper into light profile outskirts reveals an increasingly pronounced luminosity deficit in cluster QG stellar haloes relative to the field. This aligns with our findings in Section \ref{sec:results_M200}, where the lowest-mass cluster QGs in our sample ($\log M_{\star}=9.66{-}10$) exhibit less luminous haloes in more massive host clusters, likely due to increased effectiveness of environmentally-driven stripping (e.g., \citealt{Boselli2006, Peng-2010, Fang2016, Xie2025_cluster_rps}).

In conclusion, the observed trends in stellar halo assembly presented in this work (Section \ref{main-sec:results}) remain unchanged when we limit the stellar halo region to more extended radial ranges. We note that although changing the light profile integration range from $\geq2 R_e$ to $\geq3{-}5 R_e$ can significantly alter \lhaloS for individual galaxies (and even median $\mu_g$ profiles), Figure \ref{fig:discussion_MZmatch_haloregion} demonstrates that these shifts are comparable between field and cluster QG samples, leading to only minor changes in the ratio-based results across environments.

\newpage
\singlespace
\bibliography{master_bib}
\bibliographystyle{aasjournal}
\end{document}